\documentclass[lettersize,journal]{IEEEtran}

\IEEEoverridecommandlockouts

\usepackage[T1]{fontenc}
\usepackage{siunitx}
\usepackage{amsmath,amssymb,amsfonts, amsthm}
\usepackage{algorithmic}
\usepackage{graphicx}
\usepackage{textcomp}
\usepackage{xcolor}
\usepackage{caption} 
\usepackage{subcaption}
\usepackage{hyperref}
\usepackage{mathtools}
\usepackage{multirow}
\usepackage{graphicx}
\usepackage{multicol}
\usepackage{array}
\usepackage{subfiles}
\usepackage{braket}
\usepackage{url}
\usepackage{tikz}
\usetikzlibrary{automata, positioning, arrows,calc}
\usetikzlibrary{quantikz}

\newtheorem{lem}{Lemma}
\newtheorem{cor*}{Corollary}

\theoremstyle{definition}

\newtheorem*{conversionefficiency}{Up-Conversion Efficiency}

\newtheorem*{remark}{Remark}

\newcommand{\eqdef}{\stackrel{\triangle}{=}}

\begin{document}

\title{Modelling Quantum Transduction for Multipartite Entanglement Distribution}
 
\makeatletter
\newcommand{\linebreakand}{
    end{@IEEEauthorhalign}
    \hfill\mbox{}\par
    \mbox{}\hfill\begin{@IEEEauthorhalign}
}
\makeatother
    
\author{
    \IEEEauthorblockN{Laura d'Avossa, \IEEEmembership{Student~Member,~IEEE}, Angela Sara Cacciapuoti, \IEEEmembership{Senior~Member,~IEEE}, \\
    Marcello Caleffi, \IEEEmembership{Senior~Member,~IEEE}}        
    \thanks{
    The conference version \cite{DavCacCal-24} of this paper has been accepted in the Proc. of IEEE QCNC’24.}
    \thanks{The authors are with the \href{www.quantuminternet.it}{www.QuantumInternet.it} research group, University of Naples Federico II, Naples, 80125 Italy.}
    \thanks{This work has been funded by the European Union under Horizon Europe ERC-CoG grant QNattyNet, n.101169850. Views and opinions expressed are however those of the author(s) only and do not necessarily reflect those of the European Union or the European Research Council Executive Agency. Neither the European Union nor the granting authority can be held responsible for them. The work has been also partially supported by PNRR MUR RESTART-PE00000001.
     }
}
\maketitle

\begin{abstract}
Superconducting and photonic technologies are envisioned to play a key role in the Quantum Internet. However the hybridization of these technologies requires functional quantum transducers for converting superconducting qubits, exploited in quantum computation, into ``flying'' qubits, able to propagate through the network (and vice-versa). In this paper, quantum transduction is theoretically investigated for a key functionality of the Quantum Internet, namely, multipartite entanglement distribution. Different communication models for quantum transduction are provided, in order to make the entanglement distribution possible. The proposed models departs from the large heterogeneity of hardware solutions available in literature, abstracting from the particulars of the specific solutions with a communication engineering perspective. Then, a performance analysis of the proposed models is conducted through key communication metrics, such as quantum capacity and entanglement generation probability. The analysis reveals that -- although the considered communication metrics depend on transduction hardware parameters for all the proposed models -- the particulars of the considered transduction paradigm play a relevant role in the overall entanglement distribution performance.
\end{abstract}


\section{Introduction}
\label{sec:1}

Multipartite entanglement has recently recognized as a crucial resource for enabling astonishing functionalities in the Quantum Internet \cite{DurVidCir-00, IllCalMan-22,CacVisIll-23}. Thus, multipartite entanglement distribution is a key research area from a communication and network engineering perspective. Nevertheless, what is missing is a consolidated literature that treats multipartite entanglement distribution with an outlook on \textit{quantum transduction}, a crucial challenge towards the vision of the Quantum Internet. Indeed, the final stage of the Quantum Internet is envisioned as the hybridization of different quantum technologies \cite{CacCalTaf-20} -- such as \textit{superconducting} and \textit{photonic} technologies -- aiming at exploiting the complementary features of each technology and combining their different strengths to achieve superior performance and reliability.

More into details, on one hand, superconducting technology is recognized as a very promising quantum computing platform because of its capabilities to realize fast gates and its high scalability \cite{Wen-17}. However, the superconducting technology requires cryogenic temperatures, which in turn challenge the development of large-scale quantum networks. On the other hand, photonic technology is worldwide recognized as the most suitable technology for realizing the so-called \textit{flying qubits}, i.e., optical photons acting as quantum carriers, which travel along communication channels for fulfilling quantum communication needs. In fact, optical photons weakly interact with the environment (thus, less subjected to decoherence), they can be easily controlled with standard optical components and they are characterized by high-rate low-loss transmissions \cite{RenXuYon-17,CacCalTaf-20}.

However, flying qubits working at optical frequencies (typically about hundred of THz) cannot directly interact with superconducting qubits that, conversely, work at microwave frequencies (GHz). Hence, for integrating these two technologies in a quantum network, a quantum transducer is needed for converting a superconducting qubit within a network node into a flying qubit that travels through optical channels
\cite{CalCacBia-18, LauSinBar-20, LamRueSed-20, RueSedCol-16, HeaRueSah-20} and vice-versa. Yet, transduction between microwave and optical domain still represents an open problem, due to the huge frequency gap -- about five orders of magnitude -- between microwave and optical photons. 

In this paper, the transduction process is investigated in order to make the multipartite entanglement distribution possible. To this aim, we analyze the capabilities of two different paradigms underlying quantum transduction, namely \textit{direct conversion} and \textit{intrinsic entanglement generation}. Stemming from these transduction paradigms, we propose different communication models for multipartite entanglement distribution, that can be applied to any class of multipartite entangled state. We theoretically compare the proposed models through key communication metrics, such as quantum capacity and entanglement generation probability.
The conducted analysis reveals that the aforementioned communication metrics are functions of the main transducer-hardware parameters, which hugely influence the performance of the entanglement distribution.
Accordingly, we identify some crucial trade-offs in this comparison, given the current state-of-the-art technology.
The identified trade-offs are expressed in terms of achievable quantum capacity and required cooperativity (defined in Sec.~\ref{sec:3.1}), as deeply discussed in Sec.~\ref{sec:04} and  summarized in Tab.~\ref{tab:03} in that section.

The remaining part of the manuscript is organised as follows. In Section~\ref{sec:2}, we describe the problem statement by presenting two main communication paradigms for multipartite entanglement distribution. In Section~\ref{sec:3}, we develop and analyze the different communication models for multipartite distribution, by accounting for the different transducer paradigms. In Section~\ref{sec:04}, we compare the proposed models and finally, in Section~\ref{sec:5} we conclude the paper. 

\begin{figure}[t!]
    \centering
    \includegraphics[width=\linewidth]{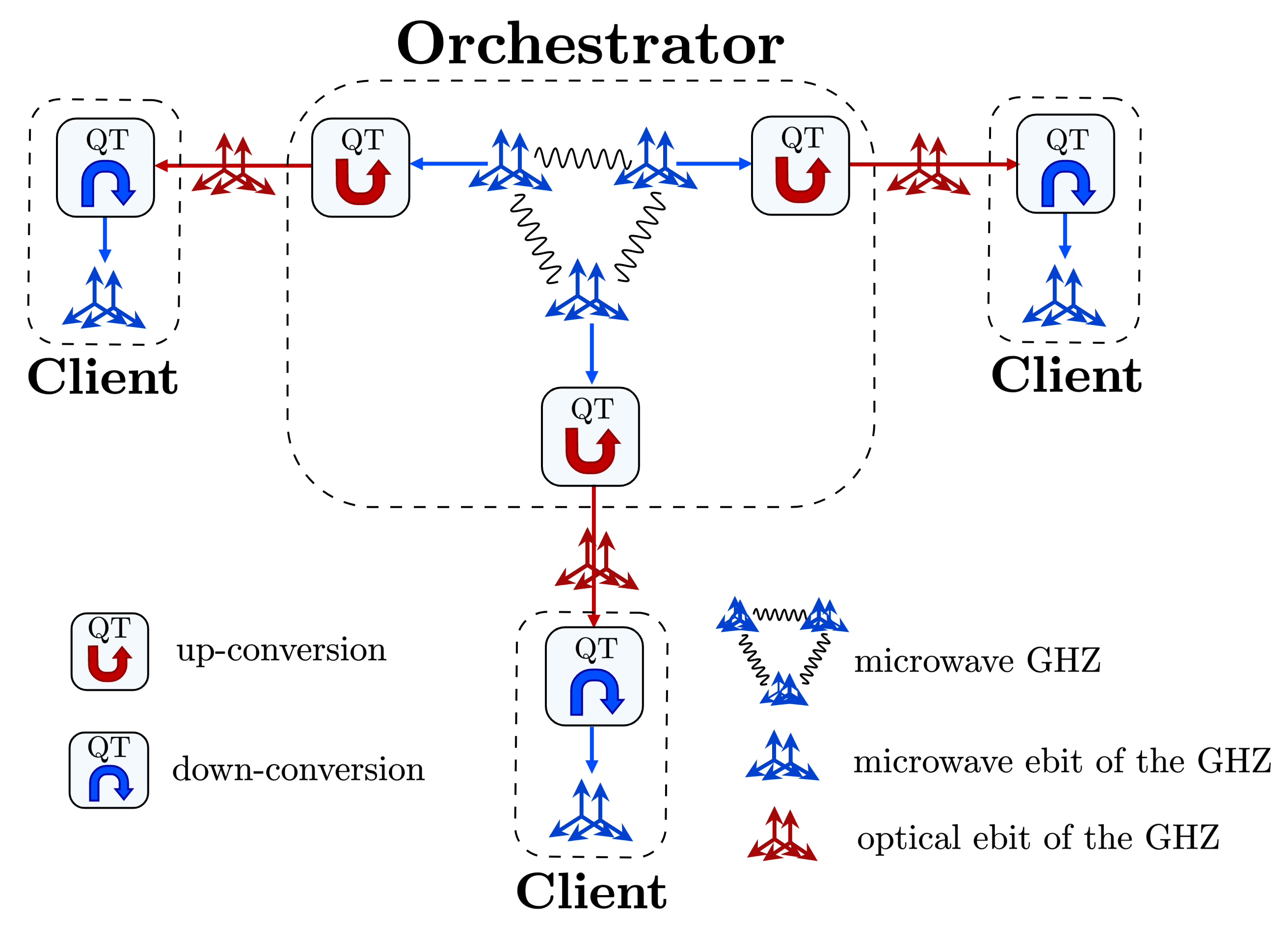}
    \caption{\textit{Direct Multipartite entanglement Distribution} (DMD) for a $3$-qubit GHZ state. The multipartite state is generated locally at the superconducting orchestrator, and it must be distributed to the superconducting nodes representing the three clients via optical quantum channels. Ebits at microwave and optical frequencies are depicted in blue and red, respectively. The Quantum Transducers (QTs) at the orchestrator realize an up-conversion of the ebits of the GHZ state, by converting them from microwave to optical frequencies. After being distributed through optical channels, the ebits are converted again into microwave frequencies with a down-conversion process implemented by the QTs at the clients.}
    \hrulefill
    \label{fig:01}
\end{figure}

\section{Problem statement: Direct vs Teleported Multipartite Entanglement Distribution}
\label{sec:2}

\begin{figure*}[t!]
	\centering
    \begin{minipage}[c]{.48\linewidth}
		\centering
		\includegraphics[width=1\columnwidth]{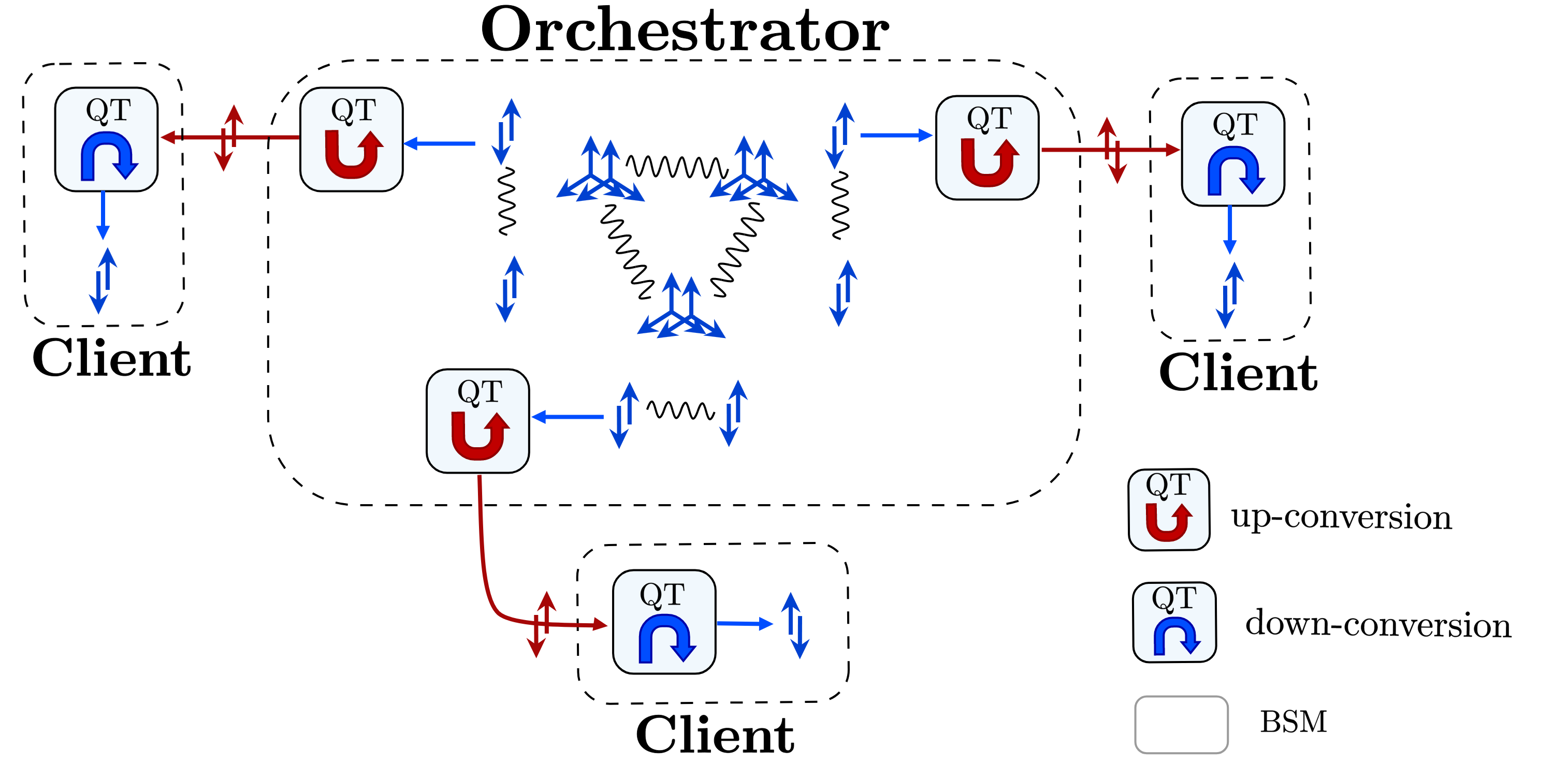}
		\subcaption{Network state before EPR pairs distribution: the orchestrator locally generates the multipartite state as well as additional microwave EPR pairs -- one for each ebit of the multipartite state that must be distributed to the clients -- that are distributed through up- and down-conversion processes, while no conversion are required for the multipartite state.}
		\label{fig:02.1}
	\end{minipage}
	\begin{minipage}[c]{.48\linewidth}
		\centering
		\includegraphics[width=1\columnwidth]{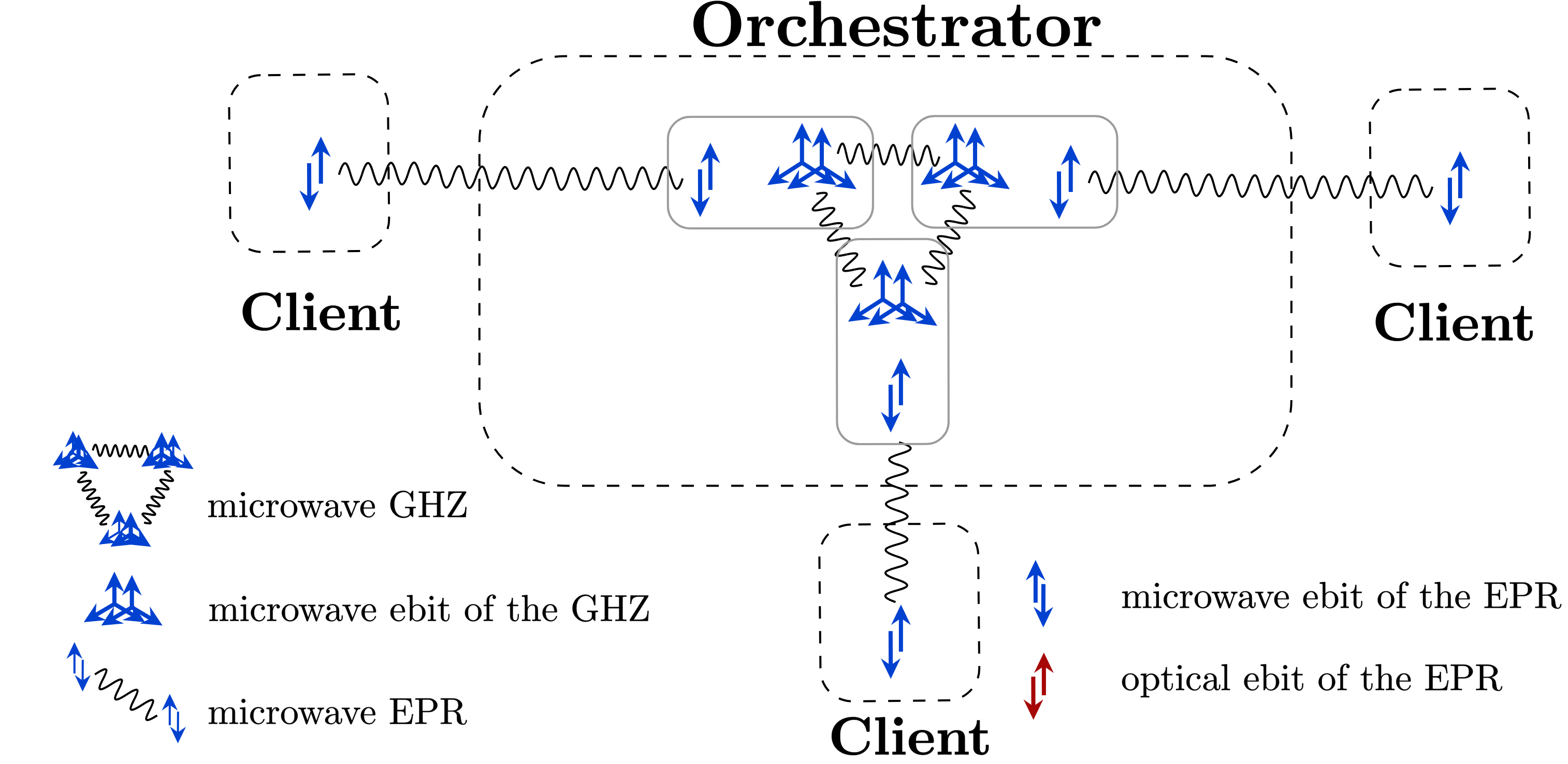}
		\subcaption{Network state after EPR pairs distribution: EPR pairs have been distributed (ideally) so that one microwave ebit is at the orchestrator and the other microwave ebit is at each client. By consuming the EPR pairs during the teleportation processes, each microwave ebit of the multipartite state -- a $3$-qubit GHZ state in this case -- is teleported at the corresponding client}
		\label{fig:02.2}
	\end{minipage}

	\caption{TMD: \textit{Teleported Multipartite entanglement Distribution} for the same multipartite state considered in Fig.~\ref{fig:01}. First, the orchestrator generates and share three EPR pairs with the clients, which represent the communication resource utilized for teleporting the multipartite state to the clients. Then, by performing local operations -- i.e., Bell State Measurements (BSMs) -- between the ebits of the EPR pairs at the orchestrator and the ebits of the GHZ state, followed by classical communications \cite{CacCalVan-20}, the $3$-qubit GHZ is shared between 3 clients.}
	\label{fig:02}
	\hrulefill
\end{figure*}

In a quantum network, generating and distributing entanglement constitutes a challenging task, due to the fragile nature of quantum states and their susceptibility to environmental noise and decoherence. The complexity of entanglement generation and distribution becomes even more demanding when it comes to multipartite entangled states.

Indeed, the generation of multipartite entanglement requires sophisticated and resource-intensive setups, often involving complex experimental apparatuses and precise control mechanisms. This makes pragmatic to assume a specialized super-node, in the following referred to as \textit{orchestrator}, responsible for the entanglement generation and distribution \cite{IllCalVis-23, CheIllCac-24, MazCalCac-25}. The orchestrator is connected via quantum channels to network nodes with lower capabilities of satisfying entanglement technological and hardware requirements, referred to as \textit{clients}. Accordingly, to eventually distribute a multipartite entangled state among the clients, the orchestrator first locally generates the multipartite entanglement state. Then, the entangled qubits -- \textit{ebits} in the following -- of the multipartite state are distributed to the clients according to selected distribution strategies \cite{CacVisIll-23}.

In the considered scenario, the orchestrator is a superconducting node, performing  quantum computations on multipartite entangled qubits. The rational for this design setup choice lies in its killer application, namely the distributed quantum computation \cite{CalAmoFer-24} in a quantum data center (also known in literature as quantum server farm) \cite{Spectrum}. Indeed, this constitutes a very hot research topic, which is gathering a huge attention by technology giants such as Amazon \cite{Amazon}, IBM \cite{IBM} and Cisco \cite{Cisco, HasEneTro-25, EneHasJia-25}. This attention is not surprising, since, in a short-term time-horizon, interconnecting different quantum processors with a quantum local-area network (Q-LAN) represents the only way (with current technology-readiness level) to scale the number of qubits beyond the thousands that are necessary to achieve fault-tolerant quantum computing \cite{MazCalCac-25, CalAmoFer-24}. Recently we showed in \cite{MazCalCac-25, SiyIllCac-24, MazCalCac-24, MazZhaChu-24} that, by leveraging the local operations at the orchestrator, it is possible to boost the network performance in terms of overhead and reliability with respect to the scenario without a supernode.
\\
In this context, distributing multipartite entangled states among remote superconducting quantum nodes constraints the adopted distribution strategy. Indeed, the chosen strategy must take into account
the huge frequency gap between intra- and inter-nodes frequencies -- namely, between microwave and optical frequencies -- by resorting to quantum transduction. And, we can distinguish two main categories for multipartite entanglement distribution strategies:
\begin{itemize}
    \item[i)] DMD: Direct Multipartite entanglement Distribution,
    \item[ii)] TMD: Teleported Multipartite entanglement Distribution.
\end{itemize}

In DMD, as suggested by the name, the ebits of the multipartite state are directly converted via quantum transducers (QTs) from microwave to optical frequencies and vice-versa, in order to be distributed to the clients, as shown in Fig.~\ref{fig:01}. Conversely, in TMD, the ebits of the multipartite state are never transducer, but they are teleported to the clients. The quantum teleportation of the multipartite state exploits EPR pairs that have been proactively generated and shared between orchestrator and the clients \cite{FanLinZhu-03}, as shown in Fig.~\ref{fig:02}. Thus in TMD, the ebits of the EPR pairs are the only ones subjected to the transduction process. The main differences between DMD and TMD are summarized in Tab.~\ref{tab:01}.

It is worthwhile to mention that, in the figures, we considered a $3$-qubit GHZ\footnote{\label{foot:01}A Greenberger–Horne–Zeilinger (GHZ) state -- formally, $\ket{GHZ}= \frac{1}{\sqrt{2}} \left( \ket{0}^{\otimes n}+\ket{1}^{\otimes n} \right)$ for a $n$-qubit state \cite{GreeHorShi-90} -- is a maximally entangled state, characterised by  maximally \textit{connectivity}. Indeed, a state is maximally connected if, for any two qubits, there exists a sequence of single-qubit measurements on the remaining qubits that, when performed, guarantee that the two qubits end up in a maximally entangled state \cite{RiefPol-11}. On the other hand, GHZ states exhibit minimum \textit{persistency} equal to $1$. Indeed, the persistency of a multipartite entangled state is the minimum number of qubits that need to be measured to guarantee that the resulting state is unentangled \cite{RiefPol-11,IllCalMan-22}.} state to be distributed to the clients.
This choice is made only for the sake of pictorial representation, since our modelling is not limited to this class of multipartite entanglement or to the number of clients. Indeed, the proposed strategies hold for a generic $n$-qubit multipartite state to be distributed to $n$ clients\footnote{To be more precise, the proposed analysis and strategies work also for the general case where the number of client is $m\leq n$. This results into a multipartite entangled state shared between the orchestrator and clients, i.e., the orchestrator retains some qubit at its side as in \cite{MazCalCac-25, SiyIllCac-24, MazCalCac-24, MazZhaChu-24}. However, in the following, for the sake of pictorial representation, we consider the scenario where the number of ebit of the multipartite state coincides with the number of clients.}.

\begin{table}[t!]
    \centering
    \includegraphics[width=0.5\textwidth]{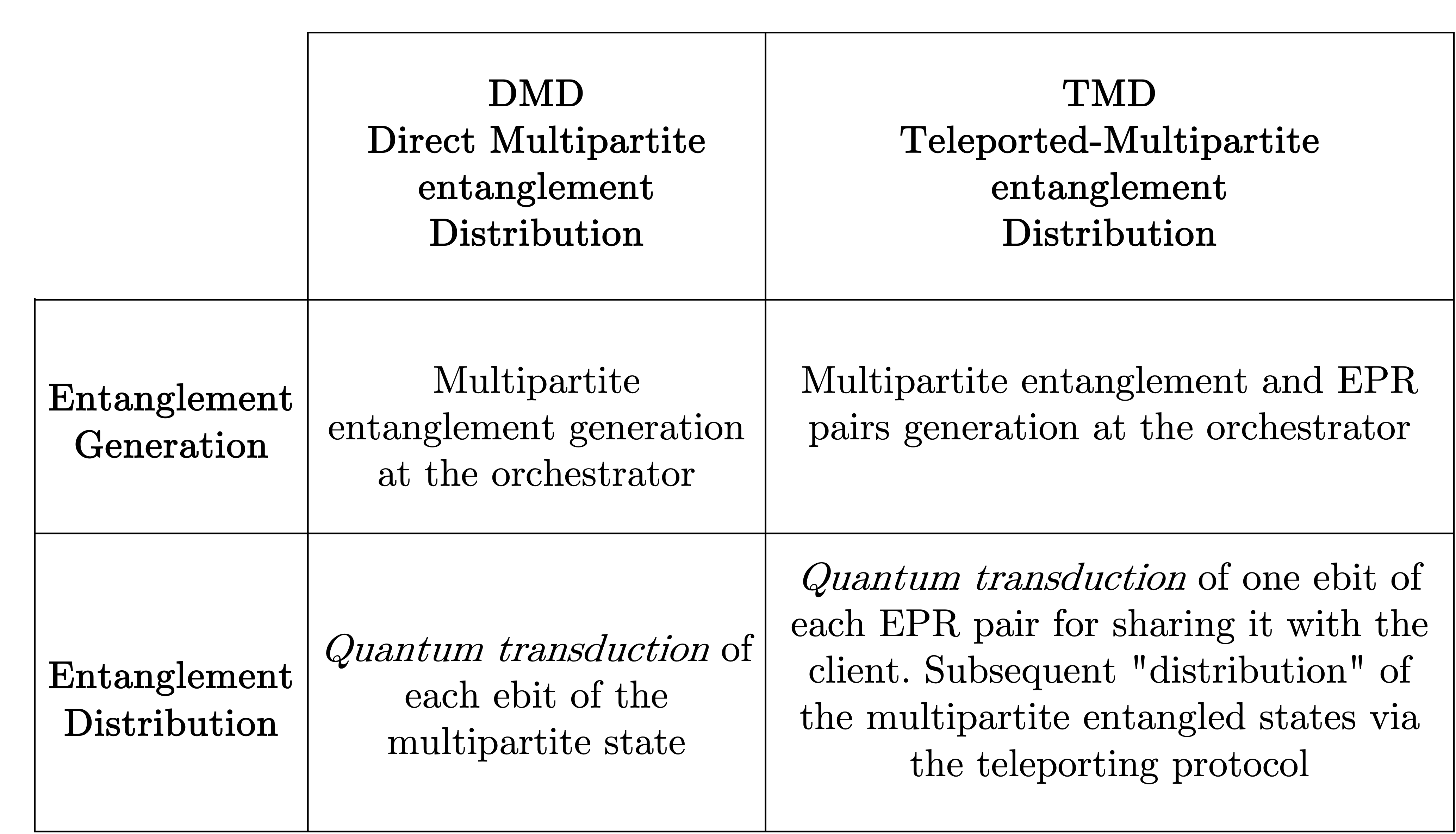} \caption{DMD versus TMD strategy for multipartite entanglement distribution.}
    \label{tab:01}
\end{table}

\subsection{DMD: Direct Multipartite entanglement Distribution}
\label{sec:2.1}

In DMD each ebit of the multipartite state can be directly distributed from the orchestrator to each client. This task requires QTs that perform two different frequency conversions for each ebit of the multipartite entangled state, so that it can be mapped from superconducting qubit to flying qubit and viceversa:
\begin{itemize}
    \item[-] \textit{up-conversion}: this process converts an ebit from microwave to optical frequencies, i.e., it converts a microwave ebit-carrier into a optical ebit-carrier,
    \item[-] \textit{down-conversion}: this process enables the inverse conversion, i.e., it converts an ebit from optical to microwave frequencies.
\end{itemize}
More into details, each ebit of the multipartite entangled state, locally generated at the orchestrator, must be distributed to remote client nodes. Unfortunately, superconducting qubits operate at microwave frequencies, which natively propagate to very limited distances, in the order of few meters \cite{KurMagRoy-18}
. Hence, interconnecting quantum nodes to distances of practical interest requires to transduce microwave frequencies into higher frequencies, such as the optical ones that allows to leverage the low-loss provided by optical fiber links. In such a way, the up-converted ebit can propagate via optical carriers to the client nodes, where it must be down-converted to microwave frequencies once again, so that it can be processed by the superconducting client.

Once this twofold up- and down-conversion process is completed for each client, the multipartite entangled state is distributed among all the nodes, as schematised in Fig.~\ref{fig:01}.

It is important to highlight that \textit{both up- and down-conversion processes are not deterministic}. As a matter of fact, there exists a non-zero probability that either or both the conversions fail \cite{HolSinZhu-20, WuCuiFan-21,Tsa-11}, with failure-probability values strictly depending on the particulars of the hardware used for implementing the microwave-optical transduction. As instance, electro-optical transduction\footnote{\label{foot:00}Electro-optical conversion exhibits several attractive features from a communication perspective with respect to other platforms -- such as transduction based on atomic-ensable or opto-electro-mechanics -- ranging from being mechanically and thermally stable through broadband to (potentially) low-noise \cite{LauSinBar-20}.} achieves values for the successful transduction probability -- also referred to as \textit{conversion efficiency} in Sec.\ref{sec:3.1} -- in the order of $10\%$ \cite{SahHeaRue-22} for both up- and down-conversion. And despite extensive research efforts in the realization of quantum transducers based on different platforms, obtaining high efficiency is still an open and crucial challenge \cite{ZhoChaHan-22}.

Accordingly, a successful DMD requires to preserve the multipartite state, originally generated at the orchestrator, during the whole distribution process. This implies that each ebit must be preserved during both up- and down-conversion processes (as well as during the carrier propagation process through the optical link) for each client. Clearly, whenever all the aforementioned processes succeed, the original multipartite state is successfully distributed among the clients. Conversely, whenever any of the mentioned processes should fails for any of the clients, then the distribution of the entire multipartite state could be compromised. This vulnerability should be avoided, as multipartite states are much more complex to generate and control with respect to bipartite entangled states. The reason for the aforementioned statement is that entanglement among the remaining ebits may or may not survive to the failure, depending on the persistence property -- defined in Footnote~\ref{foot:01} -- of the specific class of multipartite entangled state to be distributed \cite{RiefPol-11}. As an example, the direct distribution of GHZ-like states, which are characterized by the lowest persistence, requires all the ebits encoding the GHZ state to be successfully distributed to the clients in a single distribution attempt \cite{CacVisIll-23, IllCalVis-23, ZhoLiXia-23}. Hence, even the loss of a single ebit of the original GHZ state -- during up-/down-conversion or during the transmission through the optical channel -- results in the disruption of the whole multipartite state.

\subsection{TMD: Teleported Multipartite entanglement Distribution}
\label{sec:2.2}

Differently from DMD, TMD exploits the preliminary distribution of EPR pairs to the clients for eventually distributing the multipartite entangled state via quantum teleportation protocol. Specifically, once the EPR pairs\footnote{It is worthwhile to note that the generation of the EPR pairs can happen either sequentially or in parallel, depending on the characteristics of the underlying quantum technology.} have been generated, the orchestrator retains one ebit of each EPR pair while distributing the other ebit to the corresponding client. Subsequently, once the EPR ebits have been successfully received by the clients, the multipartite entangled state is teleported at the clients by performing local quantum operations (at the orchestrator, plus some correction unitaries at the clients after having received the classical bits) and classical communications (LOCC) \cite{CacCalVan-20}, as exemplified in Fig.~\ref{fig:02}.  
In the following, we assume the LOCC involved in the teleporting protocol to be noiseless. Under this hypothesis, it has been proved that the quantum teleportation is deterministic, as long as the entangled pairs distributed between the orchestrator and clients are maximally entangled \cite{FanLinZhu-03}.   
In other words, if EPR pairs are successfully generated and distributed between orchestrator and clients, the multipartite state itself is consequently successfully distributed via quantum teleportation. Clearly, entanglement purification strategies can be proactively exploited for achieving maximally entangled pairs. The rational for assuming noiseless LOCC lies in the focus of the manuscript, which aims at analysing how the transduction impacts multipartite entanglement distribution. Consequently, since in TMD strategies the transduction process acts only on the EPR pairs while the multipartite state to be distributed remains ``untouched", we neglect the noise in the LOCC that are not directly involved in the transduction problem\footnote{We refer the reader to \cite{CacCalVan-20}, for a detailed analysis about the impact of noisy LOCC on quantum teleportation in presence of superconducting nodes. In the aforementioned work, it has been showed, by exploiting the IBM quantum platform \cite{IBM_Q}, that noisy LOCC affects the fidelity of the teleported state.}.

Accordingly, TMD shifts the impact of noisy quantum transduction and noisy optical-ebit carrier propagation from multipartite ebits to  the ebits of the EPRs. Indeed, no up- or down-conversion of the multipartite ebits is required, since the multipartite state is involved only in local operations for implementing the quantum teleportation process \cite{FanLinZhu-03, ChnMarAle-19, AviRozWeh-23}, while the ebits to be transduced are those of the EPR pairs.
Consequently, i) if one of the ebits is lost (corrupted) due to the noisy transduction and/or noisy fiber transmission, only the corresponding EPR pair needs to be regenerated, rather than the entire multipartite state. This is much more easy; ii) this shifting makes \textit{TMD viable for all the classes of multipartite entanglement, regardless of their persistence properties}. 

In nutshell, TMD strategy guarantees more resilience to noise and better protection against decoherence \cite{CacVisIll-23, AviRozWen-23}, since any loss or noise would affect the EPR pair only.

For all these reasons, in the remaining part of the paper we focus on TMD only, by providing different communication models for quantum transduction in TMD and by discussing the different impact of noise on them.

\section{Quantum Transduction Models for TMD}
\label{sec:3}

As aforementioned, TMD strategy avoids to resort to quantum transduction of the multipartite ebits for coping with persistence issues and achieving better protection to noise and decoherence.

However, the ebit of each EPR pair to be shared with a client can be transduced with a cascade of up- and down-conversion as shown in Fig.~\ref{fig:02.1}, similarly to the scheme underlying DMD strategy. This scheme is referred to as \textit{vanilla quantum transduction} in the following. As we will see in Sec.~\ref{sec:3.2}, TMD with vanilla quantum transduction solves one of the two main issues exhibited by DMD, namely, its unsuitability for classes of multipartite entanglement characterized by lower-than-maximum persistence.

Yet, by utilising a cascade of up- and down-conversion TMD-\textit{vanilla} schemes do not fully overcome the severe inefficiency of direct quantum transduction, which requires a parameter regime for achieving not-null quantum capacity transduction still hard to reach with the state-of-the-art technology \cite{WuCuiFan-21, ZhoChaHan-22}. As a consequence, in the following we consider also a different communication paradigm -- referred to as \textit{intrinsic entanglement} quantum transduction -- that exploits the capabilities of the quantum transduction hardware to generate entanglement \cite{Tsa-11,RueHeaBar-19,ZhoWanZou-20, KylRauWar-23}, rather than its ability to up- or down-convert quantum states.

Before starting to model TMD for both \textit{vanilla} and \textit{intrinsic entanglement} quantum transduction, we introduce in Sec.~\ref{sec:3.1} the main parameters characterizing the performances of a quantum transducer from a communication engineering perspective.

\subsection{Quantum Transduction Efficiency}
\label{sec:3.1}

\begin{figure}[t!]
    \centering
    \includegraphics[width=0.9\columnwidth]{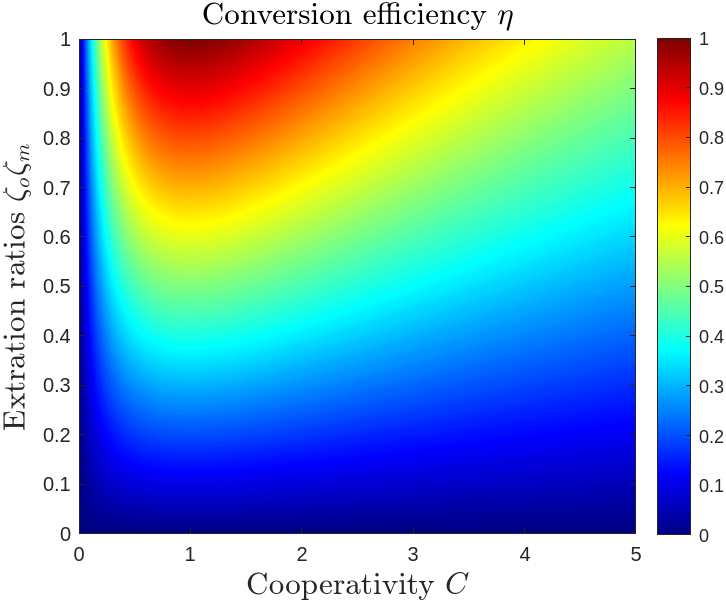}
    \caption{Conversion efficiency $\eta$ as a function of cooperativity $C$ and the product of extraction ratios $\zeta_o\zeta_m$.}
   \label{fig:03}
   \hrulefill
\end{figure}

Here we focus on \textit{electro-optical} quantum transducers due to their attractive features -- highlighted in Footnote~\ref{foot:00} -- but the theoretical analysis we develop can be easily extended to different transduction hardware, by properly accounting for the particulars in the expression of the conversion efficiency\footnote{See as instance eq. 6 in \cite{LauSinBar-20} for opto-electro-mechanics transducers.}.

In a nutshell, electro-optical transducers implement the transduction process by exploiting an input pump laser that enhances the electro-optical coupling between optical and microwave signals through the Pockels effect \cite{Tsa-10}. The main parameter governing electro-optical transduction is the conversion efficiency $\eta$, defined in the following with respect to up-conversion from microwave to optical frequencies.

\begin{conversionefficiency}
    \label{def:01}
    The up-conversion efficiency $\eta_{\uparrow}$ for converting microwave photons to optical photons via an electro-optical transducer can be derived, under resonant conditions, as \cite{Tsa-11, LiDuBaj-23}:
    \begin{align}
        \label{eq:01}
        \eta_{\uparrow} &= \frac{\gamma_{o,e} \gamma_{m,e} \, \langle n \rangle g^2 }
                {\left|\frac{ \gamma_o\gamma_m}{4} + \langle n \rangle g^2  \right|^2}
    \end{align}
    where $\gamma_{x,e}$ denotes the coupling rate for mode\footnote{With optical mode denoted with subscript $o$ whereas microwave mode denoted with subscript $m$.} $x$, $\gamma_x$ denotes the total loss rate of mode $x$, $\langle n \rangle$ denotes the average number of pump photons and, finally, $g$ denotes the electro-optic coupling coefficient. By introducing the so-called \textit{cooperativity} $C$ of the electro-optical transducer, defined as:
    \begin{equation}
        \label{eq:02}
        C= \frac{4\langle n \rangle g^2}{\gamma_o\gamma_m},
    \end{equation}
    the expression of $\eta_{\uparrow}$ in \eqref{eq:01} becomes \cite{ZhoXuCle-22}:
    \begin{equation}
        \label{eq:03}
        \eta_{\uparrow} = 4 \zeta_o \zeta_m \frac{C}{|1+C|^2}.
    \end{equation}
    In \eqref{eq:03}, $\zeta_x$ denotes the so-called extraction ratio of mode $x$, namely the ratio between the coupling rate for mode $x$ and the total loss rate of mode $x$, i.e.:
    \begin{equation}
        \label{eq:04}
        \zeta_x = \frac{\gamma_{x,e}}{\gamma_x}.
    \end{equation}
\end{conversionefficiency}

Similarly, the efficiency for down-converting optical photons into microwave photos is denoted with $\eta_{\downarrow}$. In the following, we reasonably assume the up- and down-conversion efficiencies being symmetric \cite{HeaRueSah-20, Tsa-11, FanZouCha-18}, i.e.:
    \begin{equation}
        \label{eq:05}
         \eta = \eta_{\downarrow} = \eta_{\uparrow}
    \end{equation}

From \eqref{eq:03}, it follows that high conversion efficiency $\eta$ requires both cooperativity $C$ and extraction ratios $\zeta_x$ close to 1. This is pictorially reported in Fig.~\ref{fig:03}, where the conversion efficiency $\eta$ is reported as a function of: i) cooperativity $C$, and ii) product of the extraction ratios $\zeta_o \zeta_m$.

\begin{remark}
    Regarding extraction ratios $\zeta_x$, values of $\zeta_x$ close to $1$ map into internal cavity losses $\gamma_{x,i} \eqdef \gamma_x - \gamma_{x,e}$ being relatively small \cite{Tsa-11}. It mush be noted that there is a wide-scientific consensus in considering unitary values for $\zeta_x$ feasible to achieve in the near-future. And indeed, typical values assumed in theoretical studies are around $\zeta_x=0.9$ \cite{WuCuiFan-21}, whereas experimental values in the order of $0.1-0.2$ have already been measured \cite{FuXuLiu-21}. On the contrary, unitary cooperativity is still considered beyond the state-of-the art in the near-future, with experimental values for $C$ measured recently reaching around $0.3$ \cite{SahHeaRue-22}.
    In the following subsections, we study and discuss the impact of these hardware constraints on the key functionality represented by entanglement distribution.
\end{remark}

\begin{remark}
It is evident from the previous equations, that quantum transduction is not just a merely frequency conversion process. Many factors concur to challenge the interfacing between different hardware
platforms like photonic and superconducting. For instance, in order to achieve a good transduction, the physical modes of microwave and optical systems must be matched, which includes considerations of impedance, spatial
overlap, and the temporal properties of the signals. This, as showed in \cite{RueSedCol-16}, can be captured by the electro-optic coupling coefficient $g$ in \eqref{eq:01}. Accordingly, in the following we focus on the  conversion efficiency in order to abstract from the hardware particulars. And in Sec.~\ref{sec:04}, we provide a more detailed discussion about this aspect.
\end{remark}

\subsection{TMD with ``vanilla'' quantum transduction}
\label{sec:3.2}

As mentioned at the beginning of this section, in TMD with \textit{vanilla} quantum transduction\footnote{In the following, when there is no ambiguity issue, we could refer to TMD strategy with \textit{vanilla} quantum transduction as vanilla TMD.}, one ebit of each EPR -- locally generated at the orchestrator -- is distributed to the clients with a sequence of up- and down-conversion.
Thus, we can model the TMD strategy with \textit{vanilla} quantum transduction as in Fig.~\ref{fig:02}. Specifically, $n$ microwave EPR pairs $\ket{\Phi^{o,o}_{M,M}}$ are generated at the orchestrator, where the superscript $\ket{\cdot^o}$ denotes the orchestrator as ``location'' of each ebit of the EPR and the subscript $\ket{\cdot_M}$ denotes the microwave photon domain of each ebit. Once the generation is performed, one ebit of each EPR pair generated is up-converted to optical domain, transmitted to each client and down-converted to the microwave domain therein. The process eventually results in an EPR state, denoted as $\ket{\Phi^{o,c}_{M,M}}$, since it is shared between the orchestrator $o$ and the arbitrary client $c \in \{1, \ldots, n\}$. Accordingly, once all the $n$ EPRs are distributed through the network, the overall state $\ket{\Omega^v}$ obtained with vanilla TMD before teleportation is given by:
\begin{align}
    \label{eq:06}
    \ket{\Omega^v} = \ket{\Pi} \otimes \ket{\Phi^{o,c}_{M,M}}^{\otimes n},
\end{align}
with $\ket{\Pi}$ denoting the multipartite entangled state to be distributed to the $n$ clients. The teleportation of the multipartite entangled state can be now be performed.

Given that the up and down-conversion processes are inherently probabilistic for any technology-feasible setting of cooperativity $C$ and extraction ratios $\zeta_o,\zeta_m$ in \eqref{eq:03}, it is mandatory to introduce the \textit{ebit distribution probability} as the key parameter from a communication perspective. Indeed, such a probability has to account not only for the hardware constraints but also for the propagation effects experienced by the optical photon on the quantum channel. To this aim, we adopt for the EPR distribution process the widely-used absorbing quantum channel model \cite{CacVisIll-23, Cal-17}, modelling the worst-case scenario where an entanglement carrier could not reach a client. This channel model is characterized by two elementary events: i) $E$ = ``successful distribution'', and ii) the corresponding complementary event $\overline{E}$ =``failed distribution'' representing the loss, i.e., the absorption, of the transmitted particle encoding the ebit. Furthermore, in the following, we consider as implementation of the quantum channel a telecom fiber. This is not restrictive, since the analysis in the following continues to hold by properly substituting the corresponding propagation parameter with the one describing the considered quantum channel. This choice -- i.e.,  focusing on telecom fibers -- comes from the ability, accordingly to the current state-of-the-art, of telecom fibers to offer higher communication ranges. This, in turn, allows us to focus on the quantum transduction hardware constraints, which are, from a communication perspective, less understood.

\begin{figure}[t!]
    \centering
    \includegraphics[width=0.9\columnwidth]{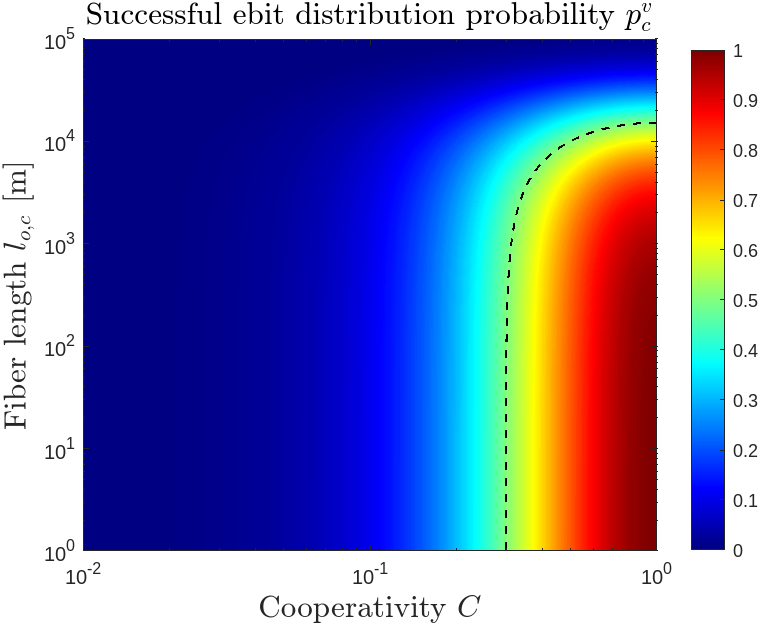}
    \caption{DMD \& vanilla-TMD: ebit distribution probability $p^v_c$ between the orchestrator and the arbitrary client $c$ as a function of cooperativity $C$ and link length $l_{o,c}$. The probability has been computed by reasonably assuming similar transduction hardware for both up- and down-conversion -- i.e., $C=C_{\uparrow}=C_{\downarrow}$ -- and by considering ideal extraction ratios -- i.e., $\zeta_o = \zeta_m = 1$. The dotted black line denotes the contour plot for $p^v_c = \frac{1}{2}$.}
   \label{fig:04}
   \hrulefill
\end{figure}

\begin{lem} \textbf{Vanilla-TMD: ebit distribution probability.}
    \label{lem:01}
    The probability $p^v_c$ of successfully distributing an ebit between the orchestrator and the arbitrary client $c$ for TMD with vanilla transduction is given by:
    \begin{equation}
        \label{eq:07}
        p^v_c = \eta^o_{\uparrow} \eta^c_{\downarrow} e^{-\frac{l_{o,c}}{L_0}}
    \end{equation}
    with $l_{o,c}$ denoting the length of the fiber link between orchestrator and client $c$, $L_0$ denoting the attenuation length of the fiber\footnote{As for today, commercial fibers feature an attenuation lower than $1$db/km. As instance, optical photons with wavelength equal to $1550$nm -- i.e., DWDM ITU $100$GHz channel number $35$ in the C band -- experience an attenuation of $0.2$dB/km, which corresponds to $L_0=22$ km \cite{SanChrder-11}.} and $\eta^o_{\uparrow}$ and $\eta^c_{\downarrow}$ denoting the efficiency of the up- and down-conversion at the orchestrator and at the client, respectively.
    \begin{IEEEproof}
        The proof follows by regarding the efficiency $\eta$ as the probability of converting an input microwave (optical) photon into an output optical (microwave) photon \cite{HeaRueSah-20, CerMahHan-18}, and by reasonably assuming up-, down-conversion and absorption as independently events.
    \end{IEEEproof}
\end{lem}

\begin{remark}
It is worthwhile to mention that hardware-agnostic parameters such as $\eta^o_{\uparrow}$ and $\eta^o_{\downarrow}$ do not provide sufficient granularity to grasp all the mechanisms/phenomena underlying a transducer process. As an example, the ``quality'' (aka, fidelity) of entanglement eventually shared between orchestrator and clients heavily depends on the type of encoding implemented within the transducer. In other words, the entanglement fidelity heavily depends on the degree-of-freedom -- e.g., polarization -- selected for encoding the entanglement within the quantum carrier (at both microwave and optical modes), as well as on the ways such a degree-of-freedom is impaired by the particular considered quantum channel. Thus, for the sake of rigour, \eqref{eq:07} provides an \textit{upper bound} for the ebit distribution probability. Stemming from the above -- and by considering that the hardware literature still lacks of a general model able to grasp all the mechanisms/phenomena underlying the electro-optical transducer process from the entanglement perspective -- we believe that it is a fair trade-off to abstract from the particulars of the considered encoding. This is even more reasonable by considering that the aim of this paper, for the first time according to the best of our knowledge, is to shed the light on a subject overlooked by the quantum communication and network communities.  
\end{remark}

\begin{remark}
    In~\eqref{eq:05}, we have reasonably assumed that the processes of up and down conversion are symmetrical. However, it is worth to highlight that the quality of the optical ebit before the down-conversion process could be smaller than the quality of microwave ebits before up-conversion,
    due to the degradation effects introduced by the optical communication channel. 
    Therefore, in Lemma~\ref{lem:01}, we modelled the decay -- exponential with the distance -- experienced by an entangled photon travelling over an optical communication channel with the term $e^{-\frac{l_{o,c}}{L_0}}$. This allows us to capture the more challenging nature of the down conversion process.
    
\end{remark}

In Fig.~\ref{fig:04}, we plot the ebit distribution probability $p^v_c$ as a function of the cooperativity $C$ and the length $l_{o,c}$ of the fiber between orchestrator and arbitrary client $c$ for both DMD and vanilla-TMD. For computing the probability, we have reasonably assumed similar transduction hardware for both up- and down-conversion, since the two conversion processes present similar hardware challenges. Specifically, we have assumed symmetric efficiencies -- i.e., $\eta^c_{\downarrow} = \eta^o_{\uparrow}$ -- characterized by identical cooperativity, i.e., $C=C_{\uparrow}=C_{\downarrow}$. Furthermore, we have considered ideal extraction ratios, i.e., $\zeta_o = \zeta_m = 1$. This assumption is not restrictive since, as pointed out in the previous subsection, it is commonly adopted in literature and experimental values with almost similar order of magnitude have been measured. Clearly, the results in Fig.~\ref{fig:04} can be easily adapted to non-ideal extraction ratio by simply weighting the probability value with the appropriate scaling factor $(\zeta_o \zeta_m)^2$.
It may seem that the overall transduction process underlying vanilla-TMD on a single orchestrator-client link -- i.e., a cascade of up- and down-conversion -- coincides with the overall process of DMD  on a single orchestrator-client link. 
However, the fundamental difference is that the former acts on the ebit of an EPR pair whereas the latter acts on the ebit of the multipartite entangled state. This implies that the probability given in \eqref{eq:07} can be used as-is for modelling the probability of successful distribution of one EPR pair via TMD and of a single ebit of the multipartite state via DMD. However, in DMD such a probability does not capture the persistency properties of the considered multipatite entanglement class. In other words, in DMD the successful distribution of an ebit on a certain orchestrator-client link does not assure the preservation of the overall multipartite entangled state, and thus it does not assure the success of the overall transduction process. As a consequence, the successful distribution of an ebit on a certain orchestrator-client link is only a necessary condition for the success the overall transduction process in DMD. In this light, in Lemma~\ref{lem:02}, we provide the necessary condition for having in DMD a non-null quantum capacity on a given orchestrator-client link.

\begin{lem} \textbf{DMD: operative region for a single orchestrator-client link.}
    \label{lem:02}
    For DMD, a necessary condition for a non-null one-way quantum capacity on a given orchestrator-client link is having $p^v_c > \frac{1}{2}$.
    \begin{IEEEproof}
         Please refer to Appendix~\ref{ap:01}
    \end{IEEEproof}
\end{lem}

Let us better clarify the rationale of this result, before discussing its implications. When it comes to DMD, the orchestrator aims at directly distributing ebits forming the overall multipartite quantum state. And, as aforementioned, the loss of even only one of these ebits can result in the loss of the whole multipartite state due to the persistence issues. Accordingly, the ebit distributed via DMD must be regarded as a quantum state to be distributed -- if possible, un-altered -- from the orchestrator to the client. Accordingly as highlighted in \cite{BennDiVSmo-97}, the one-way capacity is the right metric to adopt. Differently in TMD-based schemes \cite{ZhoChaHan-22, KylRauWar-23}, the two-way quantum capacity \cite{BennDiVSmo-97} should be considered, since the ebits of the EPR pairs -- distributed for eventually teleporting the overall multipatite state to the clients -- can be regenerated and re-distributed in case of losses, without any altering or corruption og the original multipartite state. 
Finally, we note that the conditions on $p^v_c$ provided in Lemma~\ref{lem:02} are necessary not only for the highlighted persistency constraints. Specifically, for the reasons described in the Remark after Lemma~1, direct conversions at the corresponding functional blocks and absence of absorption do not imply that the overall distributed pair shares some entanglement. This depends on other particulars underlying the considered encoding implemented within the transducer.

From Lemma~\ref{lem:02}, it results that for having unitary capacity on a given orchestrator-client link, $p^v_c$ should be unitary as well. By assuming fiber length significantly shorter than the attenuation length $L_0$, this requires unitary cooperativity $C=1$ for both the up- and the down- transducers. Yet, such a value, as pointed out in Sec.~\ref{sec:3.1}, exceeds current state-of-the-art technologies.

Hence, ``noiseless'' DMD is beyond current quantum transduction capabilities, regardless of the length of the fiber connecting orchestrator with client.

One might believe that at least ``noisy'' DMD -- i.e., DMD characterised  by non-null capacity on a given orchestrator-client link --  should be feasible, by restricting the multipartite entangled state to classes characterized by high persistence. Unfortunately this is not true since, for satisfying the condition $p^v_c > \frac{1}{2}$, the cooperativity values -- by combining the effects of both up- and down-conversion -- should satisfy $C > 2 \sqrt{2} - 2 \sqrt{2 - \sqrt{2}} - 1 \simeq 0.30$. 

In Fig.~\ref{fig:04}, we denoted the values for the cooperatives $C=C_{\uparrow}=C_{\downarrow}$ enabling non-null one-way quantum capacity with the dotted black curve. 

It is evident that the above consideration becomes even more genuine if we take into account that, for a successful distribution of the multipartite entangled state via DMD, no orchestrator-client link can fail. Thus, for having an overall non-null capacity in DMD, the condition $(p^{v})^{n} > \frac{1}{2}$ must be satisfied, with $n$ denoting the number of clients.

Differently, by considering TMD with vanilla transduction, we have the following result.

\begin{lem} \textbf{vanilla-TMD: operative region.}
    \label{lem:03}
    For TMD with vanilla transduction, a no-null two-way quantum capacity can be assured for any non-null value of $p^v_c$.
    \begin{IEEEproof}
     Please refer to Appendix~\ref{ap:02}.
    \end{IEEEproof}
\end{lem}

\begin{figure}[t!]
	\centering
	\begin{minipage}[c]{0.8\linewidth}
		\centering
		\includegraphics[width=1\columnwidth]{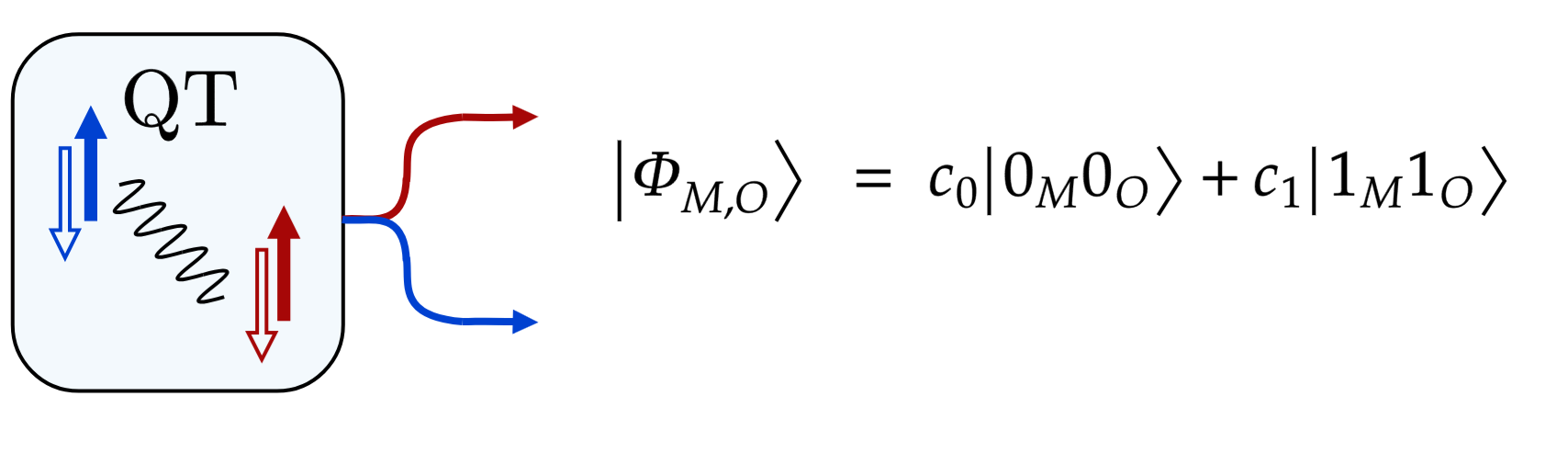}
		\subcaption{}
		\label{fig:05.1}
	\end{minipage}
	\begin{minipage}[c]{0.8\linewidth}
		\centering
		\includegraphics[width=1\columnwidth]{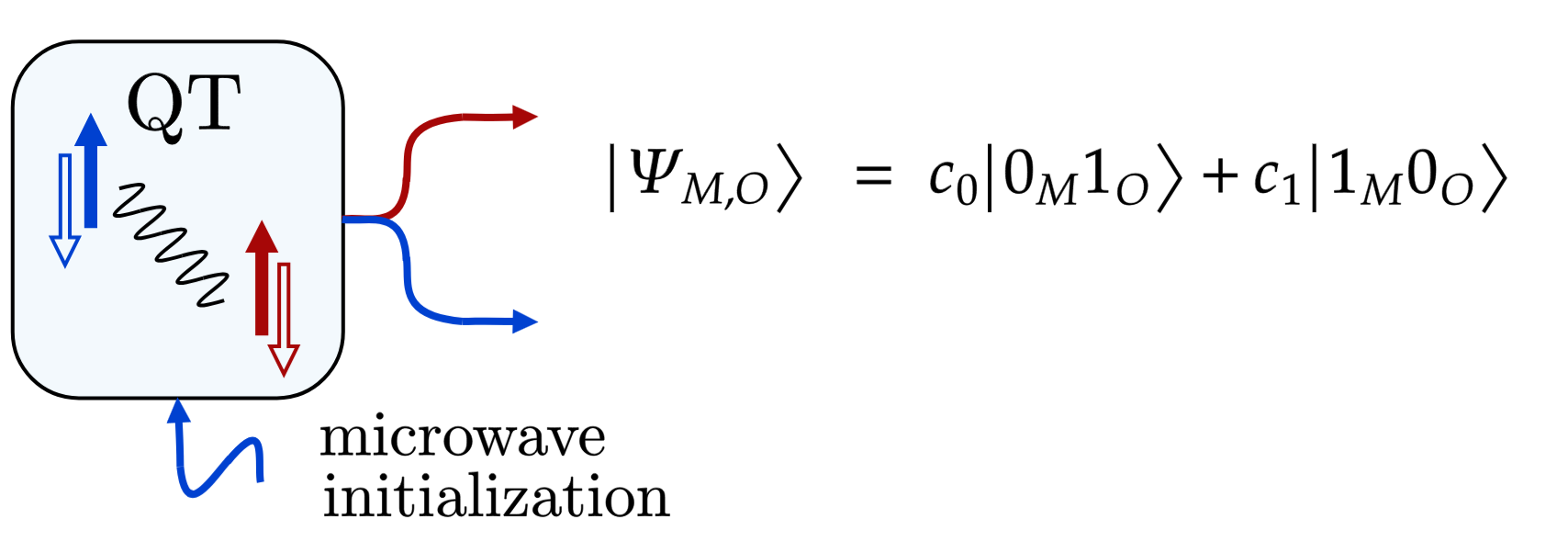}
		\subcaption{}
		\label{fig:05.2}
	\end{minipage}
	\caption{Intrinsic entanglement generation in a quantum transducer: (a) without, and (b) with initialization of the microwave field inside the cavity. Blue (red) ``up''-solid arrows represent the presence of a microwave (optical) photon, while blue (red) ``down''-empty arrows denote the absence of a microwave (optical) photon.}
	\label{fig:05}
	\hrulefill
\end{figure}

The result of Lemma~\ref{lem:03} induces us to discuss some key aspects. First, the condition in Lemma~\ref{lem:03} characterizes the entire distribution of the multipartite state, differently from Lemma~\ref{lem:02}, for the reasons highlighted above.  Second, given that the two-way quantum capacity $Q_2$ -- which, as result of Lemma~\ref{lem:03}. is equal to $Q_2=p^v_c$ -- coincides with both the one-way and the two-way entanglement distribution capacity $D_1$ and $D_2$ \cite{Wil-18,PirLauOtt-15}, it results that an average of $p^v_c>0$ ebits can be distilled from each distribution attempt, or equivalently $\frac{1}{p^v_c}$ distribution attempts must be performed in average to distil one EPR pair. Clearly, as highlighted for DMD, in these considerations we are not accounting for the effects of the encoder implemented within the transducer. If these effects were taken into account, then $p^v_c$ represents an upper bound for quantifying the distillable entanglement. From Fig.~\ref{fig:04}, it is evident that vanilla TMD does not require unitary cooperativity for achieving unitary capacities. This is remarkable with respect to DMD.

\subsection{TMD with intrinsic-entanglement quantum transduction (IE-TMD)} 
\label{sec:3.3}

As pointed out in Sec.~\ref{sec:2.1}, state-of-the-art quantum transducers suffer from very low transduction efficiency due to weak nonlinear coupling, and any attempt to enhance efficiency (by increasing the pump power \cite{HaQun-24}) would lead to an inevitable thermal noise increase due to heating.

Conversely, the parameter regime achievable with state-of-the-art technology -- when coupled with cryogenic temperatures so that thermal microwave noise can be neglected \cite{Tsa-11} -- enables the generation of \textit{intrinsic} entanglement, i.e., entanglement between two different (optical and microwave) domains \cite{WuCuiFan-21, Tsa-11}, as shown in Fig.~\ref{fig:05}. Specifically, through spontaneous parametric down-conversion \footnote{Spontaneous parametric down-conversion is a non-linear optical process where a photon spontaneously splits into two photons of lower energies \cite{Cou-18}.} (SPDC) of an input pump field, entanglement between optical and microwave fields is generated within the transducer \cite{WuCuiFan-21, RueHeaBar-19, ZhoWanZou-20, KraRanHol-21}.

In the following, we consider the entanglement generated within the QT in the form of the so-called \textit{two-photon path-entanglement} \cite{BotKokBra-00,HuvWilDow.08,MonVerCap-17}, which can be expressed with \textit{Fock} state notation as \cite{KraRanHol-21}:
\begin{align}
    \label{eq:08}
\ket{\Phi^{o,o}_{M,O}}=c_0 \ket{0_{M}^o0_{O}^o}+c_1 \ket{1_{M}^{o}1_{O}^{o}}.
\end{align}
with the subscripts $(\cdot_M)$ and $(\cdot_O)$ denoting the photon domain -- i.e., microwave or optical -- and the superscripts $(\cdot^o)$ and $(\cdot^c)$ denoting the ``location'' of each ebit, i.e., at the orchestrator or at the client. Accordingly, in \eqref{eq:08} the term $\ket{1_{M}^o 1_{O}^o}$ denotes the event in which SPDC successfully generated a couple of photons, one at microwave and the other at optical frequency. Conversely, the term $\ket{0_{M}^o 0_{O}^o}$ denotes the event in which SPDC failed, and no microwave nor optical photon have been generated. 
The coefficients $c_0$ and $c_1$ depends on the transducer hardware parameters.
Entanglement in the form of \eqref{eq:08} is generated whenever the quantum transducer is initialized with no input -- namely, no photon to be converted as in Fig.~\ref{fig:05.1} -- and the input pump field is set to operate on a frequency that is the sum of the frequencies of the optical and microwave photons (aka ``blue detuning'').

\begin{remark}
    Two assumptions underlying equation \eqref{eq:08} must be better discussed. Regarding the assumption of generating a two-level state -- i.e., the assumption of restricting the admissible system state to 2-level Fock states for each field rather than an uncountable levels -- it is not restrictive, since it simply requires a specific inizialization 
    of the microwave field inside the cavity as in Fig.~\ref{fig:05.2} \cite{KraRanHol-21}. In this case, an input pump field is set to operate on a frequency
    that is the difference of the frequencies of the optical and microwave
    photons (aka “red detuning”). The pump is exploited to obtain an entangled state in the form of 
    $\ket{\Psi^{o,o}_{M,O}} = c_0\ket{0^o_M 1^o_O}+c_1\ket{1^o_M 0^o_O}$
    , which is equivalent to \eqref{eq:08} up to a basis change \cite{KraRanHol-21, DavZhaChu-24}. 
    
    The term $\ket{0_{M}^o 1_{O}^o}$ denotes a successful conversion of the initialized microwave field within the transducer, while the term $\ket{1_{M}^o 0_{O}^o}$ denotes a failed conversion. Again, the coefficients $c_0$ and $c_1$ depends on the transducer features.
    Depending on a careful setting of the transduction hardware parameters, it is possible to obtain an EPR state (see Appendix~\ref{ap:03}) in the form: 
    \begin{align}
    \label{eq:09}
    \ket{\Psi^{o,o}_{M,O}} = \frac{1}{\sqrt{2}}(\ket{0^o_M 1^o_O}+\ket{1^o_M 0^o_O})
\end{align}
    Obviously, any hardware mismatch from the ideal setting would impact on the purity of the generated entangled pair. And we will properly account for this mismatch with the following Lemma~\ref{lem:04}.
    Thus, in the following we will use \eqref{eq:09} for the sake of simplicity.
\end{remark}

\begin{figure}[t!]
	\centering
	\includegraphics[width=0.9\columnwidth]{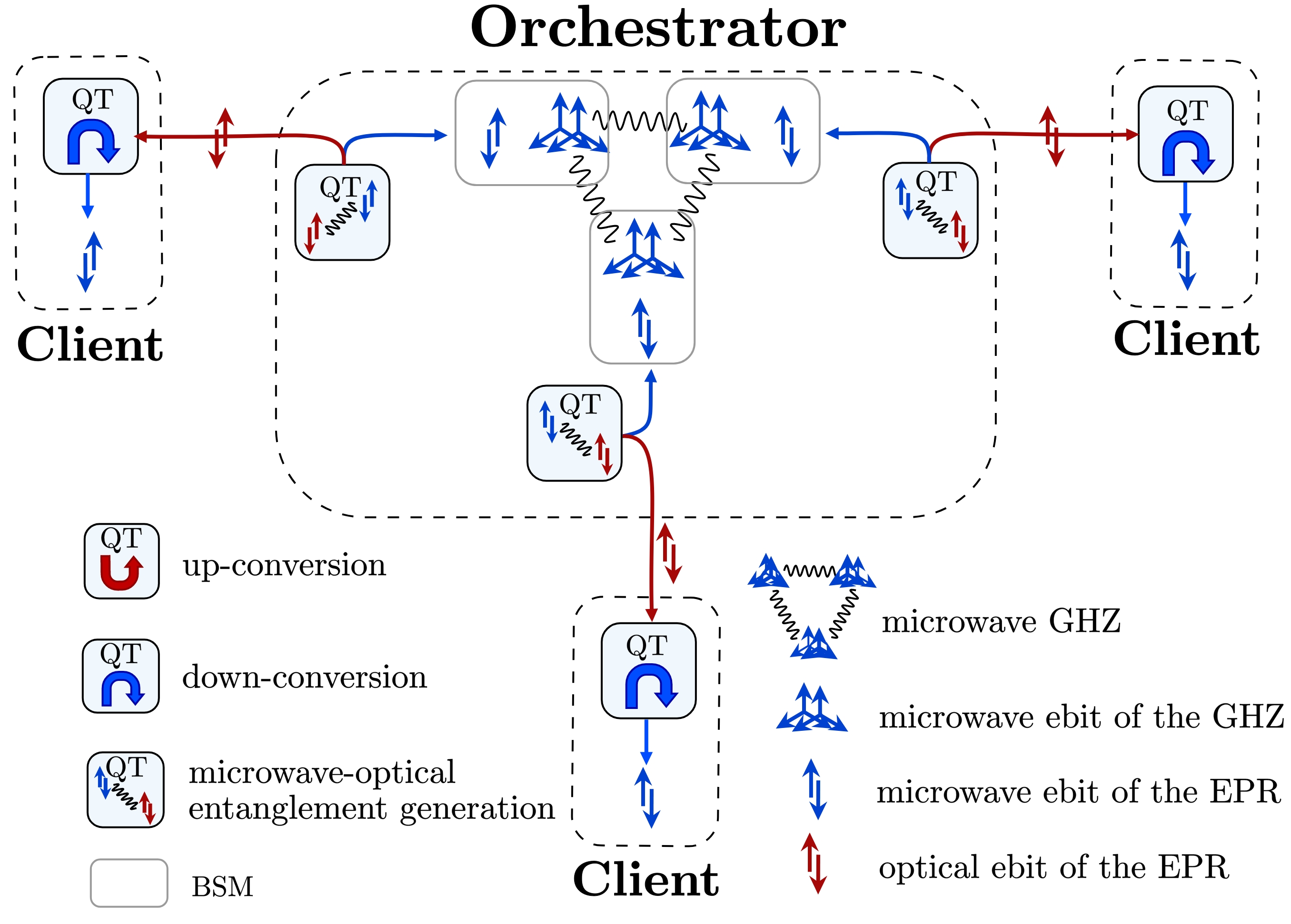}
	\caption{
   TMD with intrinsic-entanglement quantum transduction (IE-TMD) for the same multipartite state considered in Fig.~\ref{fig:01}. First, QTs generates intrinsic EPR pairs between microwave and optical domain at the orchestrator, as illustrated in Fig.~\ref{fig:05}. Ebits at microwave and optical frequencies are depicted in blue and red respectively. After being distributed through optical channels, optical ebits of the generated EPRs are down-converted to microwave frequencies by the QTs at the plain clients. Once the microwave EPRs are distributed between orchestrator and clients, the multipartite entangled state can be teleported to clients.}
    \label{fig:06}
    \hrulefill
\end{figure}

Stemming from the above, we can model the TMD strategy with intrinsic entanglement generated by the transducer at the orchestrator as in Fig.~\ref{fig:06}. In the following we refer to this strategy as \textit{TMD with intrinsic-entanglement quantum transduction }(IE-TMD). Specifically, $n$ intrinsic EPR pairs are generated at the orchestrator, with the optical ebit of each EPR pair transmitted to each client and down-converted to the microwave domain therein, resulting in the following EPR state shared between the orchestrator $o$ and the arbitrary client $c \in \{1, \ldots, n\}$:
\begin{align}
    \label{eq:10}
    \ket{\Phi^{o,c}_{M,M}} = \frac{1}{\sqrt{2}}(\ket{0_{M}^{o}0_{M}^{c}}+\ket{1_{M}^{o}1_{M}^{c}})
\end{align}
Accordingly, once all the $n$ EPR pairs are distributed through the network, the overall state $\ket{\Omega^{IE}}$ obtained with IE-TMD is equivalent to the state given in \eqref{eq:06}, and the teleportation of the multipartite entangled state can be now be performed.

The IE-TMD requires each client to be equipped with a quantum transducer capable of down-converting from optical to microwave. Thus, as vanilla-TMD discussed in Sec.~\ref{sec:3.2}, it suffers from the inefficiency of direct quantum transduction -- although limited to a single conversion (optical to micro) rather than two conversions (micro to optical and then back to micro) as for vanilla-TMD.

It is worthwhile to highlight that, similarly to up- and down-conversion, the intrinsic entanglement generation within the transducer is not deterministic, since there exists a not-null possibility that no entanglement between microwave and optical field is generated. As a matter of fact, even when some entanglement is generated, it may be non-maximally. And, indeed, we link the quality of the entanglement to the probability of generating entanglement in the following Lemma.

\begin{figure}[t!]
    \centering
    \includegraphics[width=0.9\columnwidth]{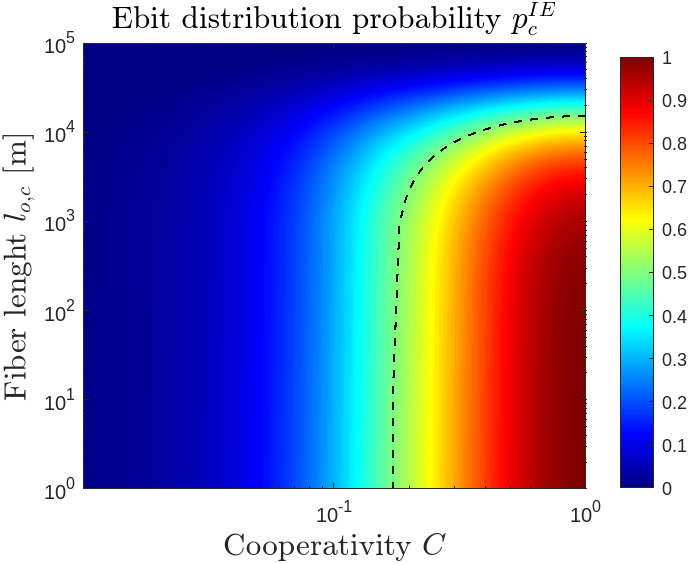}
    \caption{IE-TMD: ebit distribution probability $p^{IE}_c$ between the orchestrator and the arbitrary client $c$ as a function of cooperativity $C$ and link length $l_{o,c}$.
    The dotted black line denotes the contour plot for $p^{IE}_c = \frac{1}{2}$.}
   \label{fig:07}
   \hrulefill
\end{figure}

\begin{lem} \textbf{IE-TMD: ebit distribution probability.}
    \label{lem:04}
    The probability $p^{IE}_c$ of successfully distributing an ebit between the orchestrator and the arbitrary client $c$ for TMD with intrinsic-entanglement quantum transduction is given by:
    \begin{equation}
        \label{eq:11}
        p^{IE}_c = S \big( \eta^o_{\uparrow} \big) \eta^c_{\downarrow} e^{-\frac{l_{o,c}}{L_0}}
    \end{equation}
    where $l_{o,c}$ denotes the length of the fiber link between orchestrator and client $c$, $L_0$ denotes the attenuation length of the fiber, $\eta^o_{\uparrow}$ denotes the efficiency of the intrinsic entanglement transducer at the orchestrator, $\eta^c_{\downarrow}$ denotes the efficiency of the down-conversion at the client, and $S(\cdot)$\footnote{With a small abuse of notation, we have indicated in the argument of the Von Neuman entropy the eigenvalue determining its value rather than -- as usually done -- the density matrix on which the entropy is evaluated.} denotes Von Neuman entropy given by:
    \begin{equation}
        \label{eq:12}
        S(\eta^o_{\uparrow}) = - \eta^o_{\uparrow} \log_2(\eta^o_{\uparrow}) - (1-\eta^o_{\uparrow})\log_2 (1-\eta^o_{\uparrow})
    \end{equation}
    \begin{IEEEproof}
     Please refer to Appendix~\ref{ap:03}.
     \end{IEEEproof}
\end{lem}

\begin{figure}
	\centering
	\includegraphics[width=\columnwidth]{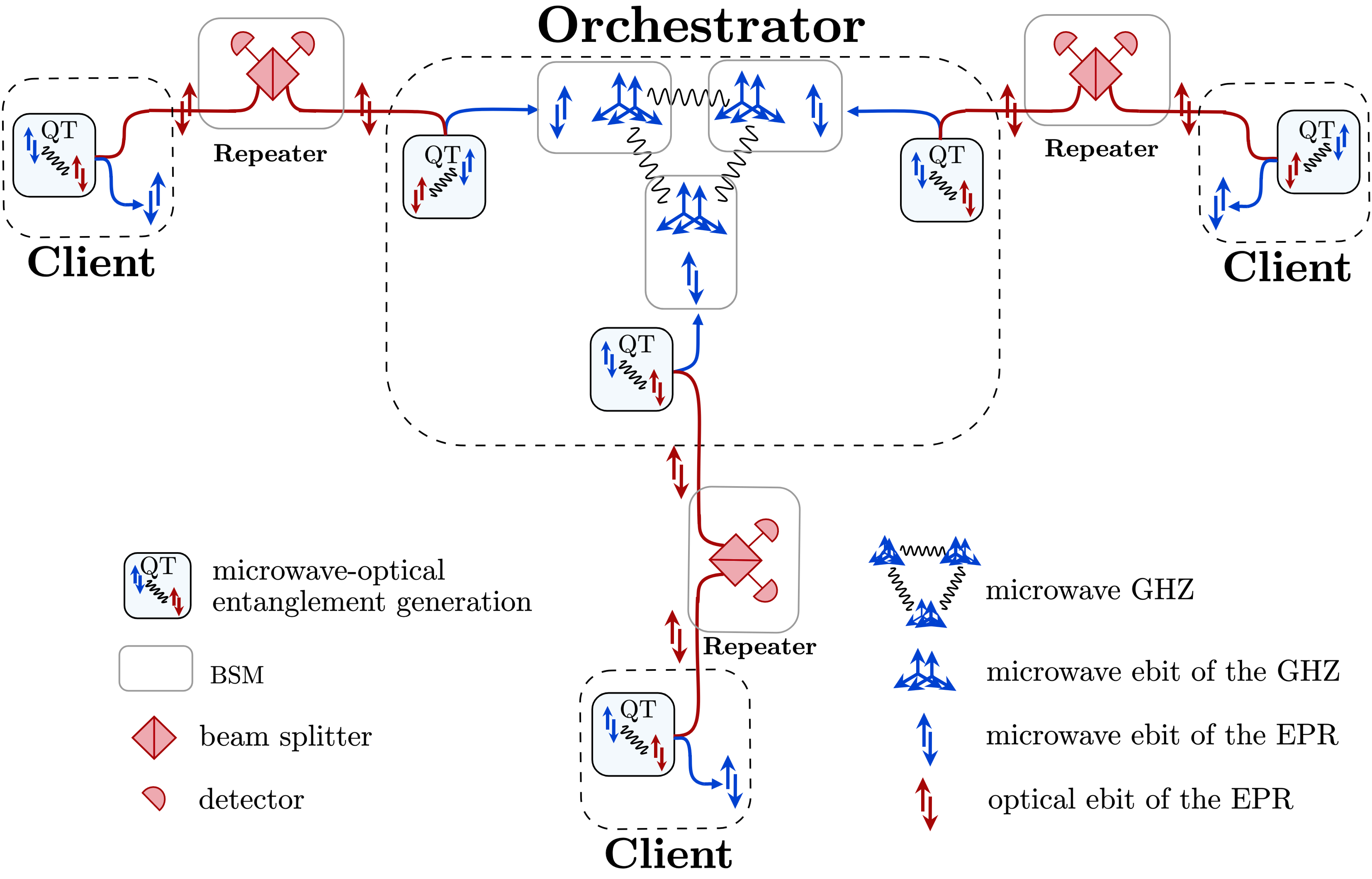}
	\caption{TMD with intrinsic-entanglement quantum transduction \& swapping (IES-TMD) for the same multipartite state considered in Fig.~\ref{fig:01}. First, QTs generates intrinsic EPR pairs between microwave and optical domain at both the orchestrator and the clients. Ebits at microwave and optical frequencies are depicted in blue and red respectively. Optical repeaters implement entanglement swapping on optical ebits so that the microwave EPRs at orchestrator and clients become entangled, and the multipartite entangled state can be finally teleported.}
	\label{fig:08}
	\hrulefill
\end{figure}

Before discussing the implication of Lemma~\ref{lem:04}, as highlighted in the appendix it is worthwhile to clarify that -- although the physical mechanisms are different -- $\eta^o_{\uparrow}$ in \eqref{eq:03} still governs the efficiency of the intrinsic
transducer at the orchestrator, since the hardware is the same. But the different physical mechanics imply less stringent requirements on the cooperativity, as discussed in the following.
In Fig.~\ref{fig:07}, we plot the successful ebit distribution probability $p^{IE}_c$ given in \eqref{eq:11} as a function of the cooperativity $C$ and the length $l_{o,c}$ of the fiber between orchestrator and the arbitrary client $c$. As mentioned above, for computing the probability, we have reasonably assumed similar transduction hardware for both up- and down-conversion, since the two conversion processes present similar hardware challenges. Yet, as discussed in the proof of Lemma~\ref{lem:04}, there exists a subtle difference in the desired operative parameters between the transducer at the orchestrator and that one at the client. Specifically, while at the client we target a unitary conversion efficiency $\eta^c_{\downarrow}$, at the orchestrator we aim to obtain a conversion efficiency $\eta^o_{\uparrow}$ equal to $\frac{1}{2}$ -- which corresponds to $C = C_{th} = 3 - 2 \sqrt{2}$ -- so that an EPR pair can be generated. From \eqref{eq:03}, we have that $\eta^o_{\uparrow}$ monotonically increases with cooperativity $C$, as visually confirmed by Fig.~\ref{fig:03}. However, as discussed in Sec~\ref{sec:3.1}, the challenge is toward obtaining higher values of $C$ rather than lowers. Hence, we avoid foolish parameter setting by computing $\eta^o_{\uparrow}$ as a function of $\min\{C, C_{th}\}$. Clearly, this choice -- reasonable from a communication perspective aiming at optimizing the entanglement distribution process -- maps into having orchestrator and client characterized by different hardware parameter setting. In this context, it is worth noting that -- differently from what would be desirable -- transducers at clients requires higher (hence, harder to achieve) cooperativity values than transducers at the orchestrator. Furthermore, in agreement with Sec.~\ref{sec:3.2}, we have considered ideal extraction ratios, i.e., $\zeta_o = \zeta_m = 1$.
\begin{figure}[t!]
    \centering
    \includegraphics[width=0.9\columnwidth]{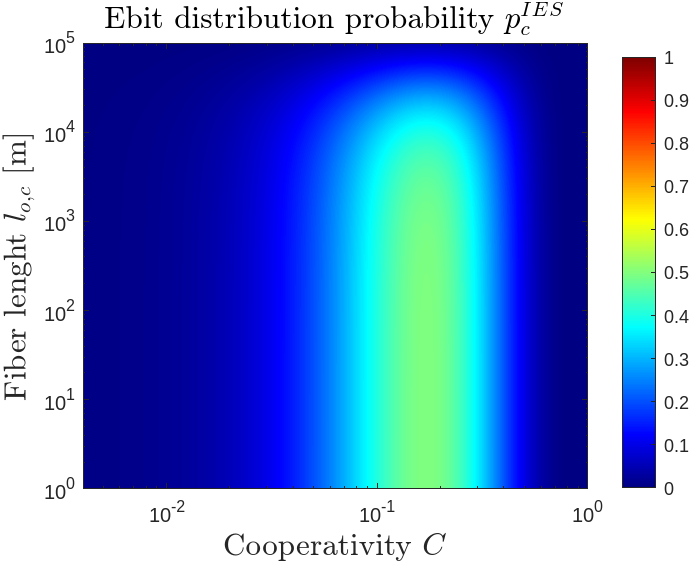}
    \caption{IES-TMD: ebit distribution probability $p^{IES}_c$ between the orchestrator and the arbitrary client $c$ as a function of cooperativity $C$ and link length $l_{o,c}$. The probability has been computed by reasonably assuming similar transduction hardware for both up- and down-conversion -- i.e., $C=C_{\uparrow}=C_{\downarrow}$ -- and by considering ideal extraction ratios -- i.e., $\zeta_o = \zeta_m = 1$.
    }
   \label{fig:09}
   \hrulefill
\end{figure}

By comparing Fig.~\ref{fig:07} with Fig.~\ref{fig:04}, we can note that the adoption of intrinsic entanglement generation at the orchestrator has improved the overall performances of the entanglement distribution. This statement becomes more clear if, by reasoning as in Lemma~\ref{lem:03}, we consider that the two-way quantum capacity for IE-TMD is equal to $p^{IE}_c$. However, by comparing Fig.~\ref{fig:07} with Fig.~\ref{fig:04}, it is also evident that the values of cooperativity $C$ enabling non-negligible entanglement distribution probability remains way beyond current state-of-the-art. This issue is overcamed by the scheme discussed in the following subsection.

\subsection{TMD with intrinsic-entanglement quantum transduction \& swapping (IES-TMD)}
\label{sec:3.4}

An additional model for TMD with intrinsic-entanglement quantum transduction is obtained by relaxing the assumption of concentrating intrinsic entanglement generation at the orchestrator, hence assuming that also clients are able to generate intrinsic EPR pairs between microwave and optical domains, and assuming that additional quantum hardware is available along the optical links.

Accordingly, the generation of the EPR pairs occurs ``at both points'' rather than at ``source only'' \cite{CacCalVan-20,KozWehVan-22}, as shown in Fig.~\ref{fig:08}. Specifically, $n$ intrinsic EPR pairs are generated at the orchestrator while $n$ intrinsic EPR pairs are generated at the clients as well. As for the IE-TMD scheme, each EPR pair is hybrid, involving both the microwave and optical domains. EPR pairs are eventually distributed between orchestrator and clients through a procedure resembling entanglement swapping \cite{BrieDurCir-98}. This scheme, referred to as \textit{TMD with intrinsic-entanglement quantum transduction and swapping (IES-TMD)}, accounts for the nature of the Fock states in \eqref{eq:09} \cite{PakZanTav-17}, and it is described in details in the following \footnote{As mentioned in the Remark of Sec.~\ref{sec:3.3}, as a seek of clarity we consider the beam splitter interaction for the hybrin EPR pair generation within the transducer hardware.}.

The Fock state generated at each client $c \in \{1, \ldots, n\}$ can be written as follows:
\begin{equation}
    \label{eq:13}
    \ket{\Psi^{c,c}_{M,O}}=\frac{1}{\sqrt{2}} \big( \ket{0_{M}^c 1_{O}^c}+ \ket{1_{M}^c0_{O}^c} \big)
\end{equation}
and $n$ Fock states as in \eqref{eq:09} are generated as well at the orchestrator.

The optical ebits of the EPR pairs are thus transmitted through optical quantum channels and reach $n$ beam splitters, one for each client, with each beam splitter followed by two detectors. The overall setup is unable to  distinguishing the \textit{which-path} information \cite{KraRanHol-21, PakZanTav-17}. A click of one of the two detectors of each setup denotes the presence of an optical photon. However, due to the path-erasure -- i.e., the impossibility of knowing whether the optical photon responsible for the detector-click has been generated at the orchestrator or at the client -- it is impossible to distinguish where the 
up-conversion process has been successful (namely, whether at the orchestrator or at the client), and thus it is impossible to distinguish whether a microwave photon is present at orchestrator or at client. This results into the generation of another form of \textit{path-entanglement} \cite{MontVivCap-15} between the microwave photons at the orchestrator and at the client \cite{KraRanHol-21}. Thus, the overall effect of beam splitter and detectors is reminiscent of entanglement swapping, since they project the received optical photons into a Bell state and the heralded signal -- i.e., the detector-click -- indicates the distribution of entangled pairs in the remote superconducting processors \cite{DuaLukCir-01} in the form of:
\begin{equation}
    \label{eq:14}
    {\ket{\Psi_{M,M}^{o,c}}=\frac{1}{\sqrt{2}}\big( \ket{0_{M}^o1_{M}^c}+\ket{1_{M}^o0_{M}^c} \big)}
\end{equation}

\begin{figure}[t!]
    \centering
    \includegraphics[width=0.9\columnwidth]{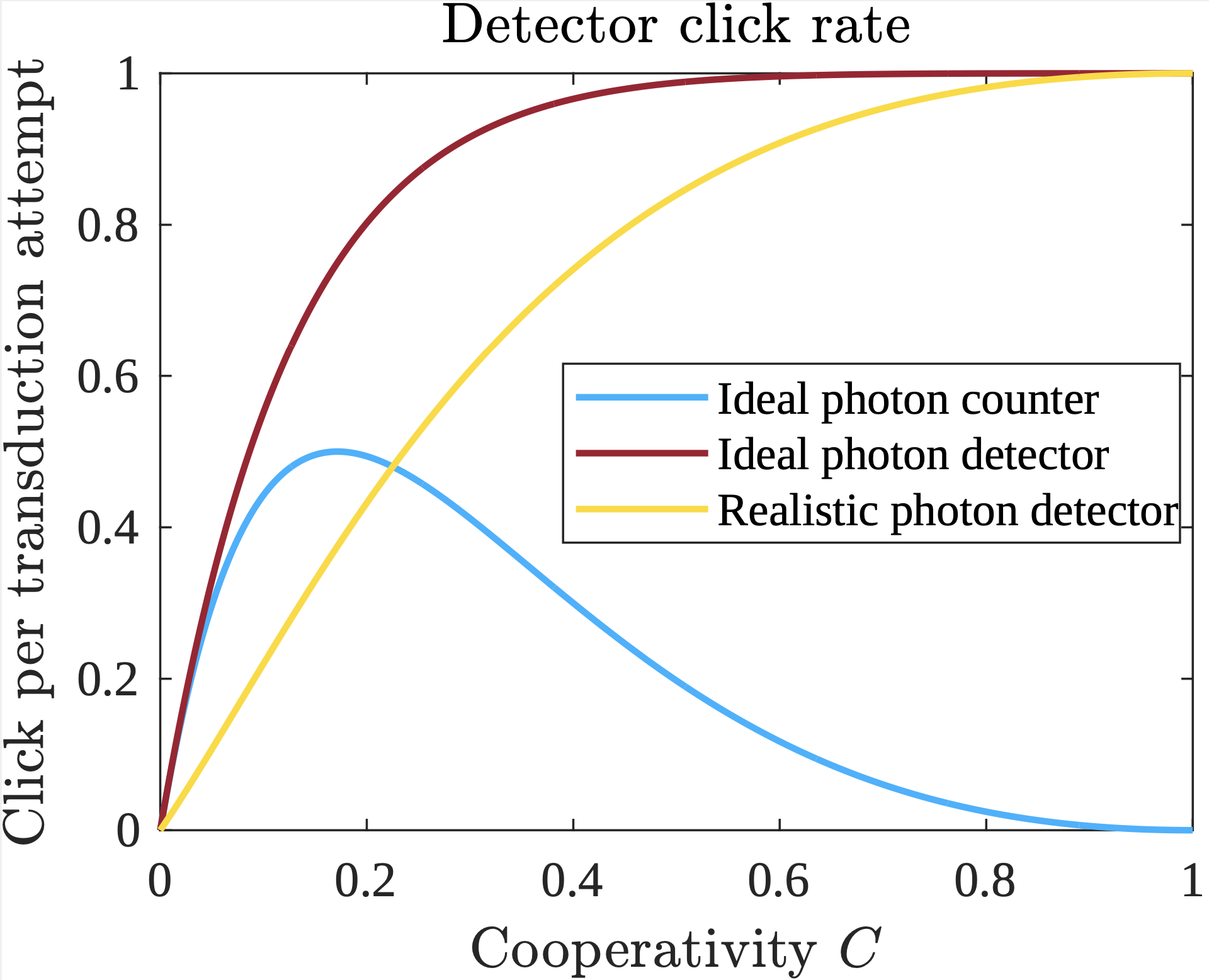}
    \caption{IES-TMD: click probability as a function of the cooperativity C for different hardware settings.}
   \label{fig:10}
   \hrulefill
\end{figure}

Accordingly, the overall state $\ket{\Omega^{IES}}$, after the swapping-like distribution procedure is once again equivalent to the state given in \eqref{eq:06}, and the teleportation of the multipartite entangled state can be performed.

\begin{remark}
It is important to highlight that, when both transducers generate optical photons, only one detector click is triggered due to path erasure.  
However, in this case, the state shared between the orchestrator and the client is $\ket{0^o_{M}}\ket{0^c_{M}}$, which is definitely not an entangled state as in \eqref{eq:14}.
However, if we reasonably assume the availability of photon-number-resolved detectors (PNRD), then it is possible to distinguish the event of receiving two optical photons -- one for each transducer in each link -- from the event where only one optical photon is received. And the double-photon event can be discarded in favour of a new distribution attempt.
\end{remark}

We are now ready to derive the ebit distribution probability.

\begin{lem} \textbf{IES-TMD: ebit distribution probability.}
    \label{lem:05}
    The probability $p^{IES}_c$ of successfully distributing an ebit between the orchestrator and the arbitrary client $c$ for TMD with entanglement-based transduction \& swapping is given by:
    \begin{equation}
        \label{eq:15}
        p^{IES}_c = S(\eta) * 2 \big[ \eta - \eta^2 \big] e^{-\frac{l_{o,c}}{2L_0}}
    \end{equation}
    where $l_{o,c}$ denotes the length of the fiber link between orchestrator and client $c$, $L_0$ denotes the attenuation length of the fiber, $\eta$ denotes the efficiency of the intrinsic entanglement generation at both orchestrator and clients, and $S(\cdot)$ denotes the Von Neuman entropy given in \eqref{eq:12}.
    \begin{IEEEproof}
        Please refer to Appendix~\ref{ap:04}.
    \end{IEEEproof}
\end{lem}
\begin{table}[t!]
    \centering
    \includegraphics[width=0.5\textwidth]{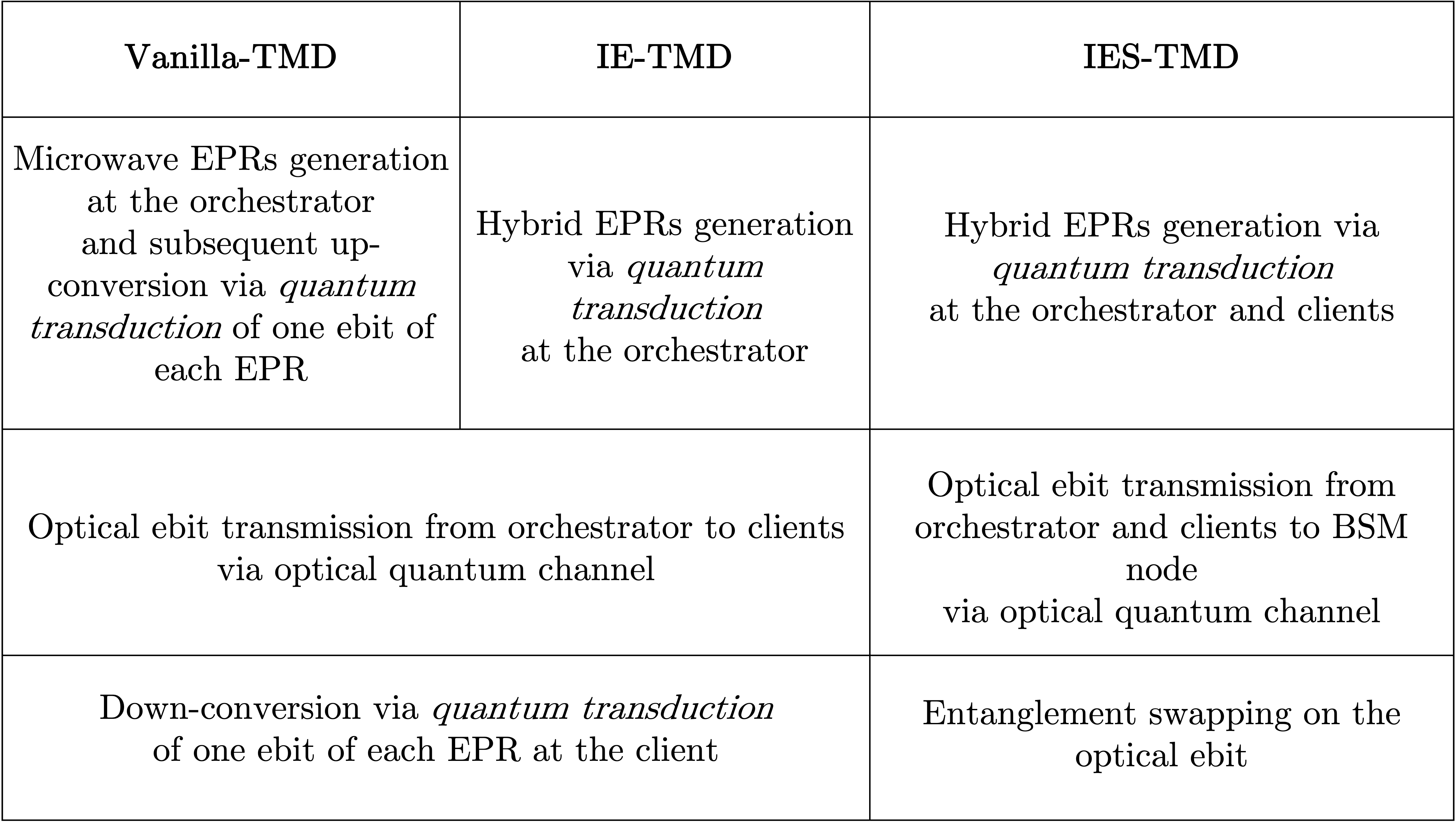} \caption{EPR pairs distribution strategies from orchestrator to clients in different TMD strategies.}
    \label{tab:02}
\end{table}
According to \eqref{eq:12}, the Von Neuman entropy is unitary -- i.e., a maximally entangled pair is generated -- when $\eta=1/2$. This, in turn, implies that, by neglecting the fiber attenuation, the maximum achievable probability $p^{IES}_c$ is $1/2$.

In Fig.~\ref{fig:09}, we plot the successful ebit distribution probability $p^{IES}_c$ given in \eqref{eq:15} as a function of the cooperativity $C$ and the length $l_{o,c}$ of the fiber between orchestrator and the arbitrary client $c$. Again, we assumed similar transduction hardware for both orchestrator and client. Comparing the simulation results of IES-TMD the ones of IE-TMD, it is evident that we have a trade-off. Indeed, in IES-TMD there is an improvement in terms of minimum cooperativity $C$ that allows a nonzero entanglement distribution probability, but this comes at the cost of a probability that never reaches 1. This, in turn, implies that the the two-way quantum capacity for IES-TMD does not reach one. This statement is more clear  if, by reasoning as in Lemma~\ref{lem:03}, we consider that $Q_2$ is equal to $p^{IES}_c$.

In summary the IES-TMD strategy allows a moderate reduction of the cooperativity $C$ requirements for enabling non-negligible entanglement distribution probability, but at the price of:
\begin{itemize}
    \item[i)] introducing additional hardware requirements in the form of the beam-splitters and detectors required for entanglement swapping, and
    \item[ii)] outperforming worse than any other strategy in therms of maximum value for the quantum capacity.
\end{itemize}

Yet, it must be acknowledged that the ``key'' advantage of such a strategy lies in the possibility of heralding entanglement via off-the-shelf hardware -- i.e., via PNRD.
Specifically, a detector click for each transduction attempt constitutes an indicator for identifying the generation of entanglement between orchestrator and client, without destroying it.

Also in Vanvilla-TMD and IE-TMD, it is possible to herald the entanglement with photon detectors by exploiting other degree of freedom such as time-bin \cite{ZhoHanTan-20, ZhoWanZou-20}.
This also implies the introduction of additional heralding setups.
On the contrary, in IES-TMD strategy entanglement heralding is embedded in the setup itself and, for this reason, it is not necessary to exploit other degree of freedom and/or to introduce any additional heralding equipment.

\begin{figure}[t!]
    \centering
    \includegraphics[width=0.9\columnwidth]{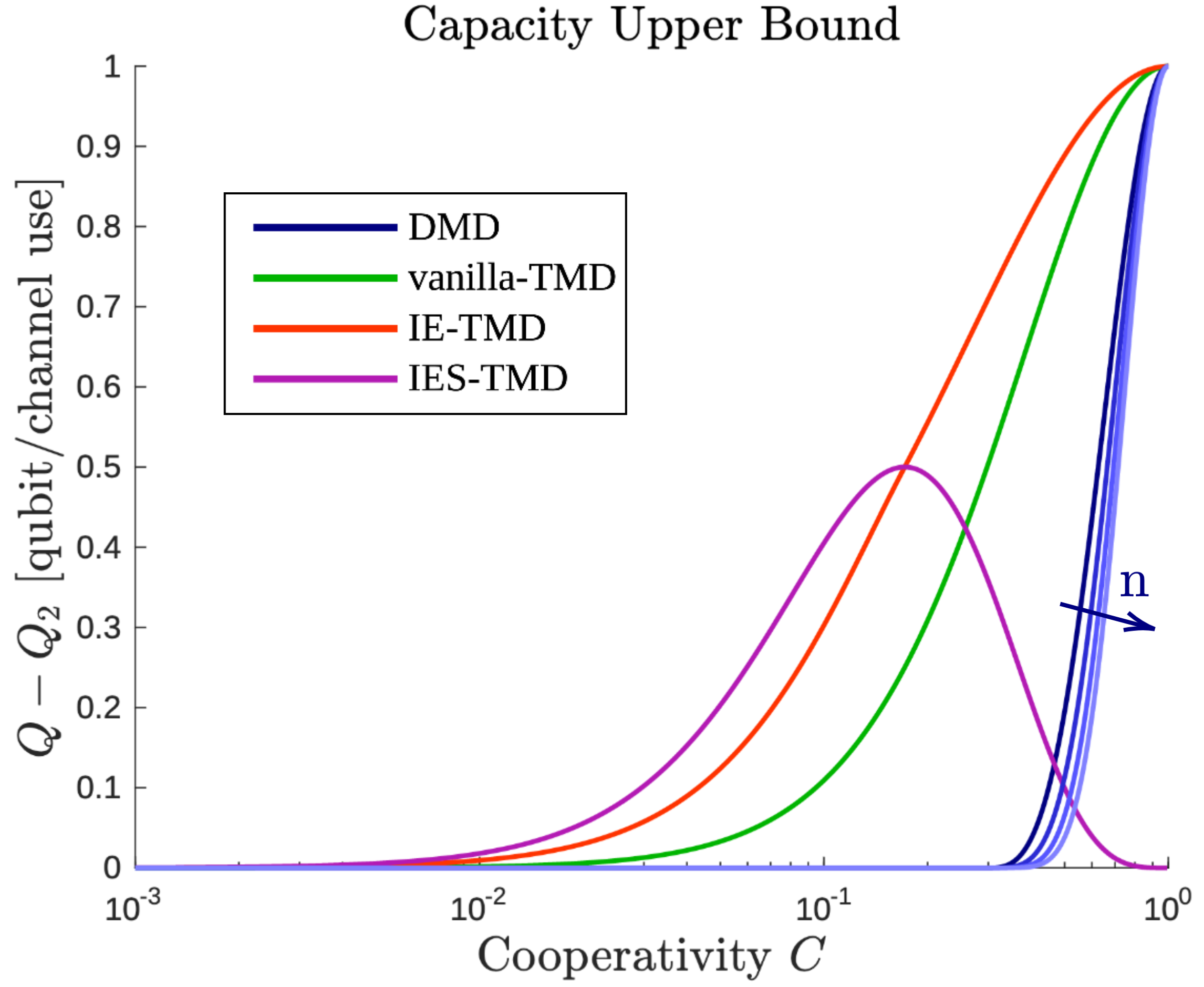}
    \caption{Capacity upper-bounds as a function of cooperativity $C$ for the different multipartite entanglement distribution strategies. Fiber effects on optical photon propagation assumed as negligible, i.e., $l_{o,c} << L_0$.}
   \label{fig:11}
   \hrulefill
\end{figure}

Furthermore, it is worth emphasizing that, in Lemma~\ref{lem:05} we considered the availability of
PNRD. If this hardware requirement cannot be satisfied, we can assume the availability of simpler single-photon detectors (SPD), which are not able to distinguish whether a click is due to one or two temporally-coincident photons. In this case, the probability of a single detector click includes also the event where both the transducers at orchestrator and client generate an optical photon, and therefore it can be expressed as follows:
\begin{align}
    \label{eq:16}
    p_c^{SPD}= 2 \eta- \eta^2
\end{align}
Both the  
PNRD probability (see \eqref{eq:29} in the Appendix~\ref{ap:04}) and the SPD probability given in \eqref{eq:16} are reported in Fig.~\ref{fig:10} as a function of the cooperativity $C$. It is easy to note that the SPD probability is significantly larger than the PNRD probability. Yet, in case of single photon detectors, only a fraction $\frac{2 \eta- 2 \eta^2}{2 \eta - \eta^2}$ of clicks correspond to entanglement generation, with the remaining click fraction corresponding to a failed attempt. We can further generalize \eqref{eq:16} by taking into account the detectors efficiency $\eta_d$, one of the main parameter to quantifying the performance of SPD \cite{Had-09}.
Specifically, if we consider IDQ single photon detector with detection efficiency $\eta_d=0.25$ \cite{Online}
 running in gated\footnote{So that we can reasonably neglect the dark-counts by synchronising the transduction attempt, i.e. the pump pulse, with the detection window.} mode \cite{Had-09}, we obtain the green curve in Fig.~\ref{fig:10}. Clearly, while both ideal- and realistic-SPD click rates tends to $1$ as the cooperativity increases, the PNRD click reaches its maximum equal to $\frac{1}{2}$ for $C=C_{th}$, as discussed above. Yet, the former click rates includes failed attempts that do not herald entanglement.

\begin{table}[t!]
    \centering
    \includegraphics[width=0.5\textwidth]{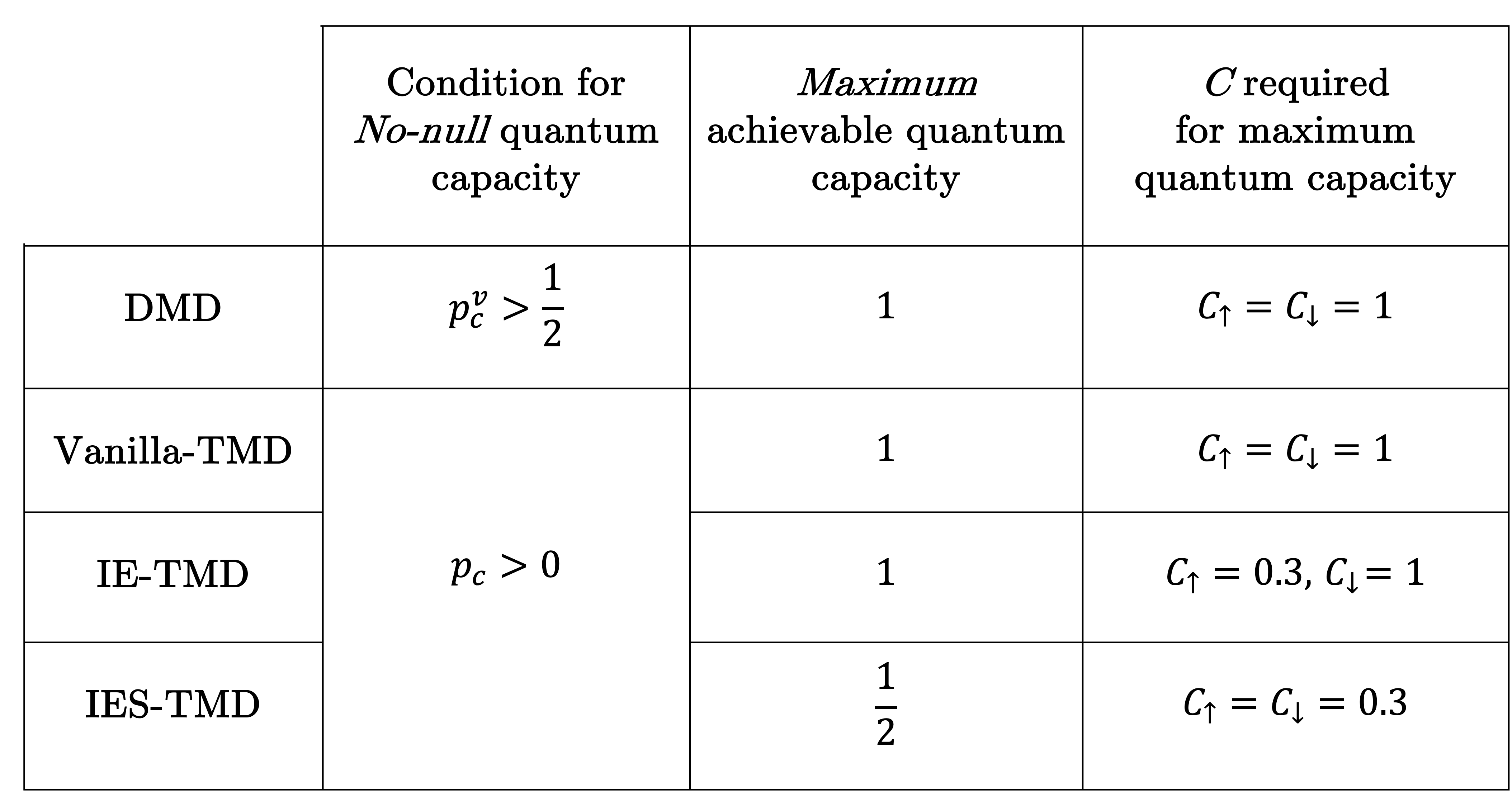} \caption{Strategies comparison in terms of achievable quantum channel capacity and the corresponding required cooperativity $C$.}
    \label{tab:03}
\end{table}

\section{Discussion}
\label{sec:04}

Stemming from the developed analysis, it appears that the nature of the entangled pairs exploited in the TMD strategies are diverse.
More into details, in the case of Vanilla-TMD, the EPR pairs, generated by the orchestrator, are at microwave frequencies. Therefore, in order to be distributed to the clients, one ebit for each EPR pair has to be up-converted to optical frequency through the transducer available at the orchestrator. On the contrary, in the case of IE-TMD and IES-TMD strategies, hybrid entangled pairs are generated: one ebit of each EPR pair is at microwave frequency and the other ebit is at optical frequency. Thus, by adopting IE-TMD or IES-TMD strategy, the EPR generation process is directly performed at the transducer hardware, by exploiting two mode squeezing or beam splitter interaction, respectively, as detailed in Sec.~\ref{sec:3.3}. The aforementioned difference in the type and nature of the exploited EPR pairs in the different TMD strategies are summarized in Tab.\ref{tab:02}.\\ 
We graphically summarise in Fig.~\ref{fig:11} the communication performances of the proposed strategies, in terms of achievable quantum capacity, as a function of the cooperativity $C$. \\
It is worthwhile to stress that the transducer efficiency does not provide sufficient granularity to grasp all the mechanisms/phenomena underlying a
transducer process, as the type of encoding implemented within the transducer. Thus, we theoretically evaluated  upper bound for the ebit distribution probabilities and thus for the quantum capacities. This is the reason for which in the legend of Fig.~\ref{fig:11} we indicated ``capacity upper bounds''.
For the sake of clarity, Tab.~\ref{tab:03} shows the strategies comparison in terms of achievable quantum capacity for the required values of cooperativity $C$.\\
The aforementioned consideration allows us to further stress an aspect highlighted in the final remark of Sec.~\ref{sec:3.1}. Specifically, as observed, quantum transduction is not just a merely frequency conversion process.  
Indeed, several factors 
impact the transducer conversion efficiency and consequently the failure conversion probability \cite{RueSedCol-16, XuFuZou-21}.  
Nevertheless, the hardware literature still lacks of a general model able
to grasp all the mechanisms/phenomena underlying the electro-optical transducer process from the entanglement generation and distribution perspective. Stemming from this, and by considering that our main goal is to start a dialogue among different communities (i.e. communication, physics, and hardware), we choose to propose a first 
model of quantum transducer in which the
conversion efficiency constitutes the main parameter for characterizing its performances. The rational for this choice is that several factors determining the conversion failure can be embedded within the
conversion efficiency. 
An example to all, to ensure a successful conversion, the mode phase-matching condition has to be fulfilled. The effect of mode mismatching is typically mitigated through filtering techniques \cite{JorFeiJim-24}, that reduces the probability of conversion and consequently the conversion efficiency.
As result, the mode-mismatch can be embedded in the conversion efficiency parameter through the parameter $g$ in \eqref{eq:01}, as shown in \cite{RueSedCol-16}.
Our design choice allows not only to abstract from the particulars of the specific technology underlying the transducer hardware, but also to easily and progressively take into account technological advancements that improve the transducer performances.\\

\section{Conclusions}
\label{sec:5}

Efficient microwave-optic quantum transduction is a key ingredient for the deployment of quantum networks, enabling the interconnection between remote superconducting nodes with optical channels. 
To this aim, in this paper we presented different communication models for multipartite entanglement distribution based on recent advances in quantum transduction devices to go beyond direct conversion.
We have showed that TMD strategies shift from ``preserving the multipartite state during frequency up- and down conversions'' to ``generating high-fidelity EPR pairs" for teleporting the multipartite state. And, indeed, among the TMD strategies we analysed the IES-TMD, which exhibits the remarkable property of requiring $50\%$ conversion efficiency for achieving effective transduction. This results in a less stringent requirement on cooperativity, but at the cost of an entanglement probability that never reaches 1. Additionally, the exclusive advantage of IES-TMD scheme over all others is the ability to herald entanglement through detector clicks.
Therefore, by switching from direct transduction to intrisic entanglement generation and performing entanglement swapping, we are able to address the stringent error probability constraints required by direct transduction. The conducted analysis is far from being exhaustive. The reason is that the hardware influences hugely the quantum transduction performance, and there is not a unique hardware solution available in literature.  Nevertheless, we provided guidelines for the QT comparison from a communication engineering perspective. And we do hope that this paper will strike up a dialogue among the different research communities involved to converge on a standard reference model, which would be of paramount importance for quantum network development.

\begin{appendices}
\section{Proof of Lemma~\ref{lem:02}}
\label{ap:01}
 
By accounting for the specificity of the different functional blocks, it results that the cascade of up-conversion, fiber channel and down-conversion can be modeled as an overall quantum erasure channel
       
        (QEC), where the incoming ebit is replaced by the erasure state $\ket{\epsilon}$ with probability $(1 - p^v_c)$. Formally, the output density matrix $\rho_o$ on a given orchestrator-client link is given by:
        \begin{equation}
            \label{eq:17}
            \rho_o = p^v_c \rho_i + \left( 1-p^v_c \right) \ket{\epsilon}\bra{\epsilon},
        \end{equation}
        with $\rho_i$ denoting the input density matrix. Accordingly, the proof follows by considering that the QEC channel exhibits a one-way quantum capacity $Q$ equal to $Q = \max\{0,1-2\epsilon\} = \max \{0, 2 p^v_c -1 \}$ \cite{BennDiVSmo-97}, with $\epsilon$ denoting the erasure probability. Thus, a non-null capacity requires $p^v_c >\frac{1}{2}$.

\section{Proof of Lemma~\ref{lem:03}}
\label{ap:02}
 By modelling the overall process as a quantum erasure channel -- as done for Lemma~\ref{lem:02} -- we have that the two-way quantum capacity $Q_2$ is given by $Q_2 = 1 - \epsilon = p^v_c$, with $\epsilon$ denoting the erasure probability \cite{BennDiVSmo-97}. Hence, for any $p^v_c>0$, we have $Q_2>0$.

\section{Proof of Lemma 4}
\label{ap:03}

To prove the Lemma we first need to prove that the efficiency $\eta^o_{\uparrow}$ in \eqref{eq:03} governs the intrinsic entanglement generation at the orchestrator between microwave and optical domain. To this aim, we assume to generate intrinsic entanglement thought the transducer initialization with a microwave photon and its probabilistic up-conversion into an optical photon \cite{KraRanHol-21, DavCalWan-23}, as discussed in the Remark of Sec.\ref{sec:3.3} (``red detuning''). This conversion leads to a \textit{beam splitter} interaction Hamiltonian\footnote{The interaction Hamiltonian is obtained in the condition of the rotating-wave approximation \cite{Tsa-11}.} $H_{int}=g\sqrt{\langle n \rangle}(ab^\dagger+a^\dagger b)$, where $a$ and $b$ are the optical and microwave mode, respectively, and $a$ and $a^\dagger$ denote the lowering and raising operator. As done in Sec.~\ref{sec:3.1}, we assume unitary extraction ratios $\zeta_o = \zeta_m = 1$, i.e., null internal cavity losses for both optical and microwave fields. To properly analyse the transducer, we exploit the input-output relations in the Heisenberg-Langevin form \cite{Tsa-11}:
\begin{align}
    \label{eq:18}
    \frac{da}{dt} &= ig\sqrt{\langle n \rangle}b - \frac{\gamma_o}{2} a + \sqrt{\gamma_{o}} a_{\text{in}} \\
    \label{eq:19}
    \frac{db}{dt} &= ig\sqrt{\langle n \rangle}^*a - \frac{\gamma_m}{2} b + \sqrt{\gamma_{m}} b_{\text{in}} \\
    \label{eq:20}
    a_{\text{out}} &= \sqrt{\gamma_{o}} a - a_{\text{in}} \\
    \label{eq:21}
    b_{\text{out}} &= \sqrt{\gamma_{m}} b - b_{\text{in}}
\end{align}
where $a_{in},b_{in}$ and $a_{out},b_{out}$ are the input and output modes of the optical and microwave field, respectively. By solving such equations utilising the Laplace transform \cite{Tsa-11}, the input-output relations\footnote{We consider a weakly coupled system, i.e., $g<<\gamma_{x}$ \cite{KraRanHol-21}. Therefore, \eqref{eq:18} is computed at zero detuning, as it represents the optimal solution in this condition \cite{ZhoChaHan-22, ZhHanHon-20}.} are:
            \begin{equation}
                \label{eq:22}
                \begin{pmatrix}
                    a_{\text{out}} \\
                    b_{\text{out}}
                \end{pmatrix}
                =
                \overbrace{\begin{pmatrix}
                        L_{o,o} & L_{o,m} \\
                        L_{m,o} & L_{m,m}
                    \end{pmatrix}}^{\text{$\eqdef L$}}
                \begin{pmatrix}
                    a_{\text{in}} \\
                    b_{\text{in}}
                    \end{pmatrix}
            \end{equation}
            where:
            \begin{equation}
            \label{eq:23}
                L= \frac{1}{d}
                \begin{pmatrix}
                    \frac{\gamma_{o}\gamma_{m}}{4} - g^2\langle n \rangle & ig \sqrt{\langle n \rangle \gamma_{o}\gamma_{m}} \\
                    ig \sqrt{\langle n \rangle ^* \gamma_{o}\gamma_{m}} & \frac{\gamma_{o}\gamma_{m}}{4} - g^2\langle n \rangle 
                \end{pmatrix},
            \end{equation}
            with $d=\frac{\gamma_{o}\gamma_{m}}{4}+g^2\langle n \rangle$. By accounting for the expression of $\eta^o_{\uparrow}$ in ~\eqref{eq:01}, it results that $|L_{o,m}|= |L_{m,o}|=\sqrt{\eta^o_{\uparrow}}$ for any value of the extraction ratios $\zeta_x$, whereas $|L_{o,o}|=|L_{m,m}|$ follows from our assumption\footnote{In general, when $\zeta_x<1$, $|L_{o,o}|\neq |L_{m,m}|$, due to the presence in \eqref{eq:23} of the parameters $\gamma_{x,e}$.} of unitary extraction ratios. 
            
            By accounting for \eqref{eq:23}, it results that matrix $L$ determining the input-output relations between the microwave and optical fields corresponds to the matrix of a \textit{lossless beam splitter} with a transmission coefficient $T$ and a reflection coefficient $R$ \cite{OuHonMan-87, CamSalTei-89}, i.e.:
            \begin{equation}
                \label{eq:24}
                \begin{pmatrix}
                    a_{\text{out}} \\
                    b_{\text{out}}
                \end{pmatrix}
                =
                \begin{pmatrix}
                    \sqrt{R} &  \pm i\sqrt{T} \\
                    \pm i\sqrt{T} & \sqrt{R}
                \end{pmatrix}
                \begin{pmatrix}
                    a_{\text{in}} \\
                    b_{\text{in}}
                \end{pmatrix},
            \end{equation}
            where $T+R=1$, $RT^*+TR^*=0$ and with $T=\eta^o_{\uparrow}$ and $R=1-\eta^o_{\uparrow}$. In terms of Fock states \cite{LaP-22}, given an input state $\ket{\psi_{in}} =b^\dagger_{in}\ket{00}=\ket{1_M0_O}$ the output state can be expressed as follows:
            \begin{align}
                \label{eq:25}
                \ket{\psi_{out}} &= \sqrt{\eta^o_{\uparrow}}\ket{0_M1_O}+\sqrt{1-\eta^o_{\uparrow}}\ket{1_M0_O}.
            \end{align}
            Whenever $\eta^o_{\uparrow}=T=1$, $\ket{\psi_{in}}$ is up-converted into $\ket{\psi_{out}}=\ket{0_M1_O}$ with $100\%$ up-conversion efficiency and the hardware behave as an ideal quantum transducer. Conversely, if $\eta^o_{\uparrow}=\frac{1}{2}$, the $L$ matrix becomes \cite{Ulf-03}:
            \begin{equation}
            \label{eq:26}
                \begin{pmatrix}
                    a_{\text{out}} \\
                    b_{\text{out}}
                \end{pmatrix}
                =\frac{1}{\sqrt{2}}
                \begin{pmatrix}
                    1 & i \\
                    i & 1
                \end{pmatrix}
                \begin{pmatrix}
                    a_{\text{in}} \\
                    b_{\text{in}}
                \end{pmatrix}
            \end{equation}
            and, in this condition the transducer acts as a \textit{50/50 lossless beam-splitter} \cite{Ulf-03} generating a Bell State $\ket{\psi_{out}} = \frac{1}{\sqrt{2}}(\ket{0^o_M 1^o_O}+\ket{1^o_M 0^o_O})$. 
            From \eqref{eq:25}, for any setting of operative parameters resulting in $\eta^o_{\uparrow} = \frac{1}{2}$ -- i.e., $C = C_{th} = 3 - 2 \sqrt{2}$ -- an EPR pair is generated. Conversely, whenever $C \neq C_{th}$, we have two cases. For degenerate values of $\eta^o_{\uparrow}$ -- i.e., $\eta^o_{\uparrow}=0$ or $1$ -- the transducer outputs a single mode and no entanglement is generated, whereas for non-degenerated values of $\eta^o_{\uparrow}$ an odd superposition of states $\ket{01}$ and $\ket{10}$ -- which is not a maximally entangled state -- is generated. 
  
  For quantify the entanglement generation probability, we resort to the concept of \textit{entanglement of distillation} \cite{nielsen00}, which is an upper bound of the number $n$ of EPR pairs that can be distilled from $m$ copies of a pure state using LOCC. Formally, for a pure state with density matrix $\rho$, the distillable entanglement $E(\rho)$ is given by \cite{HorHorHor-98}:
            \begin{equation}
                \label{eq:27}
                E(\rho) = S(\rho_A) \eqdef -Tr(\rho_A \log_2 \rho_A)
            \end{equation}
            where $S(\rho_A)$ denotes the Von Neuman entropy and $\rho_A$ denotes the reduced density matrix of either of its reduced states. By accounting for \eqref{eq:25}, we have that:
            \begin{equation}
                \label{eq:28}
                \rho_A = \begin{pmatrix}
                    \eta^o_{\uparrow} & 0 \\
                    0 & 1 - \eta^o_{\uparrow}
                \end{pmatrix},
            \end{equation}
and thus \eqref{eq:27} coincides with \eqref{eq:12}.
The overall proof follows by accounting for \eqref{eq:27}, by regarding $\eta^c_{\downarrow}$ as the probability of converting an input optical photon into an output microwave photon at the client and by accounting for the length of the fiber between the orchestrator and the client. 

\section{Proof of Lemma 5}
\label{ap:04}
By accounting that $\eta$ models the probability of generating an optical photon at either the orchestrator or the client, and by fairly assuming such events as independent, the probability of generating an optical photon in one (but not both the) transducer is:
        \begin{align}
            \label{eq:29}
            \eta_{\uparrow}^o (1-\eta_{\uparrow}^c) +(1-\eta_{\uparrow}^o) \eta_{\uparrow}^c = 2 (\eta-\eta^2)
        \end{align}
        where we assumed same conversion efficiency for both the transducers, i.e., $\eta_{\uparrow}^c = \eta_{\uparrow}^o = \eta_{\uparrow}$. By reasoning as in Lemma~\ref{lem:04}, we quantify the probability through the entanglement of distillation. Hence, by accounting for \eqref{eq:27} and \eqref{eq:29}, and by assuming that beam splitter and detectors in the mid-point for each fiber link, \eqref{eq:15} follows.
\end{appendices}

\vskip -2\baselineskip plus -1fil

\begin{IEEEbiography}
[{\includegraphics[width=1in,height=1.25in,clip,keepaspectratio]{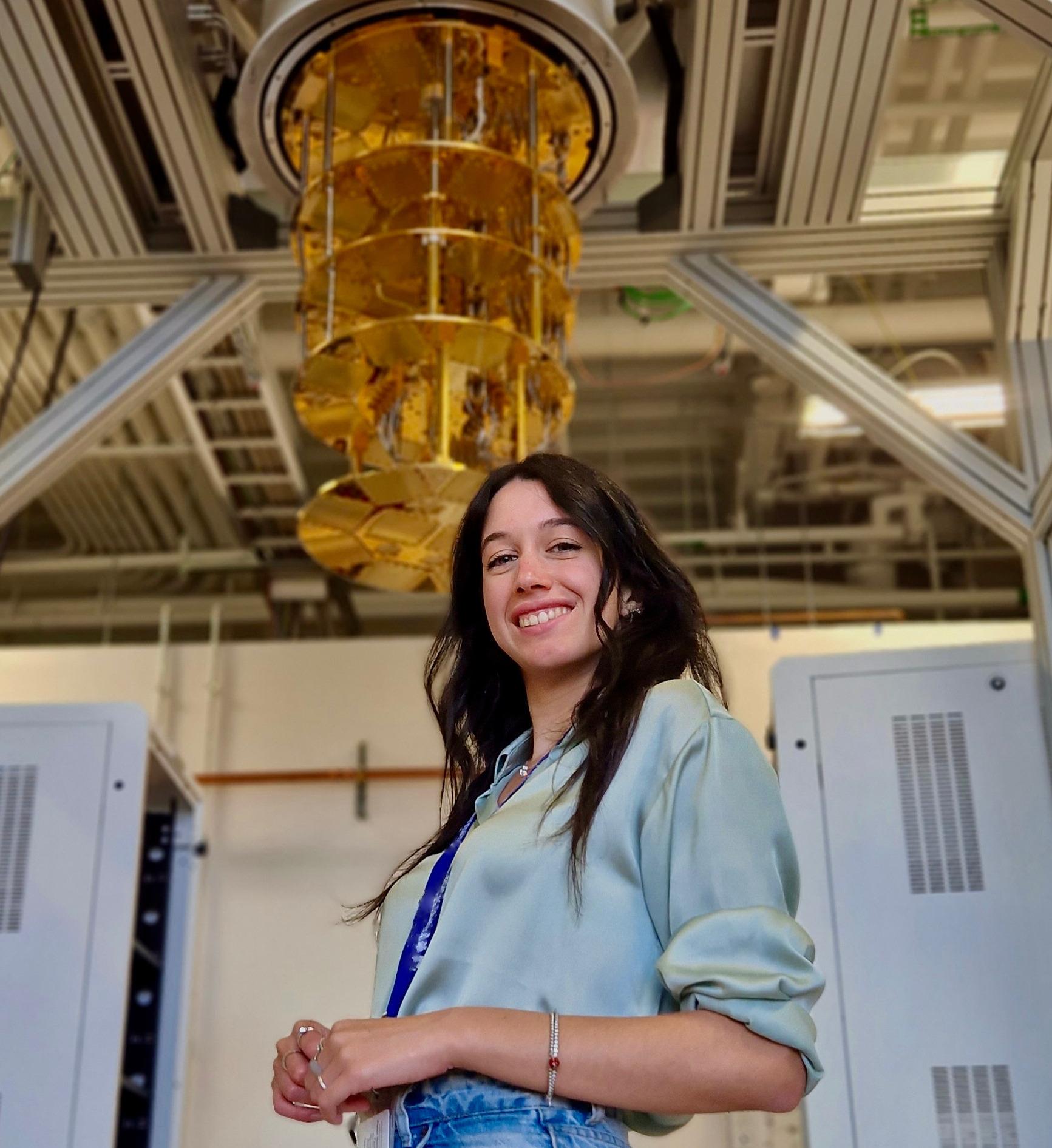}}]{Laura d'Avossa} \, (Student Member,IEEE) received the M.Sc degree in Electrical Engineering in 2022 (summa cum laude) from University of Naples Federico II (Italy). Since 2022 she is a member of Quantum Internet Research Group, FLY: Future Communications Laboratory. In 2023, she was an Intern at Fermi National Accelerator Laboratory and in 2024 she was a research assistant at Argonne National Laboratory, USA.
Currently, she is a Ph.D. student within the Quantum Technologies doctoral program at the University of Naples Federico II. She is also a member Co-chair of N2Women.

Her research interests involve quantum communications and quantum transduction.
\end{IEEEbiography}
 
\begin{IEEEbiography}
[{\includegraphics[width=1in,height=1.25in,clip,keepaspectratio]{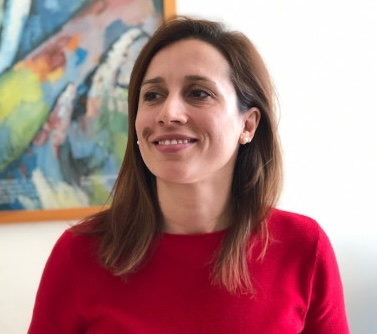}}]{Angela Sara Cacciapuoti} \, (Senior Member,
IEEE) is a Professor of Quantum Communications and Networks at the University of Naples Federico II (Italy). She is the PI of the \textit{ERC-CoG 2024 n.101169850 ``QNattyNet''}.
Her work has appeared in first tier IEEE journals and she received different awards, including the ``2024 IEEE ComSoc Award for Advances in Communication'', the ``2022 IEEE ComSoc Best Tutorial Paper Award'', the ``2022 WICE Outstanding Achievement Award'' for her contributions in the quantum communication and network fields, and ``2021 N2Women: Stars in Networking and Communications''. Lately, she also received the IEEE ComSoc Distinguished Service Award for EMEA 2023, assigned for the outstanding service to IEEE ComSoc in the EMEA Region. She also served as IEEE ComSoc Distinguished Lecturer, with lecture topics on the Quantum Internet design and Quantum Communications. Currently, she serves as Member of the TC on SPCOM within the IEEE Signal Processing Society. Moreover, she serves as Area Editor for IEEE Trans. on Communications and as Editor/Associate Editor for the journals: npj Quantum Information, IEEE Trans. on Quantum Engineering, IEEE Communications Surveys \& Tutorials. She served as Area Editor for IEEE Communications Letters (2019 - 2023), and she was the recipient of the 2017 Exemplary Editor Award of the IEEE Communications Letters. In 2023, she also served as Lead Guest Editor for IEEE JSAC special issue ''The Quantum Internet: Principles, Protocols, and Architectures''. From 2020 to 2021, Angela Sara was the Vice-Chair of the IEEE ComSoc Women in Communications Engineering. Previously, she has been appointed as Publicity Chair of WICE. From 2017 to 2020, she has been the Treasurer of the IEEE Women in Engineering (WIE) Affinity Group of the IEEE Italy Section. Her research interests are in Quantum Information Processing, Quantum Communications and Quantum Networks.
\end{IEEEbiography}

\begin{IEEEbiography}
[{\includegraphics[width=1in,height=1.25in,clip,keepaspectratio]{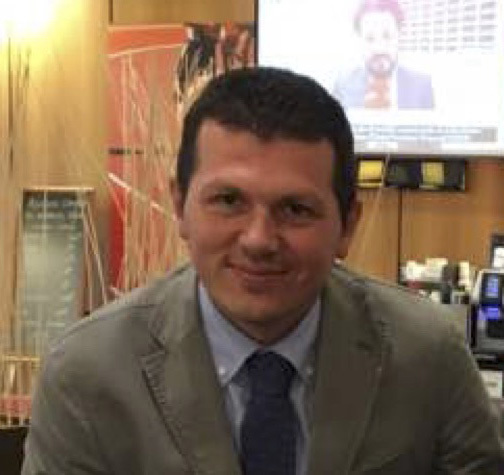}}]{Marcello Caleffi} \, (Senior Member,
IEEE) is currently a Professor of advanced quantum networks with the DIETI Department, University of Naples Federico II, Naples, Italy, where he co-lead the \textit{Quantum Internet Research Group}. His work appeared in several premier IEEE Transactions and Journals, and he was the recipient of multiple awards, including the \textit{2024 IEEE Communications Society Award for Advances in Communication} and the \textit{2022 IEEE Communications Society Best Tutorial Paper Award}. He is currently the Editor/Associate Editor of IEEE Transactions on Wireless Communications, IEEE Transactions on Communications, IEEE Transactions on Quantum Engineering, IEEE Open Journal of the Communications Society, and IEEE Internet Computing. He has been Chair/TPC Chair for several premier IEEE conferences. In 2017, he has been appointed as a \textit{Distinguished Visitor Speaker} from the IEEE Computer Society and he has been elected treasurer of the IEEE ComSoc/VT Italy Chapter. In 2019, he has been also appointed as a \textit{member of the IEEE New Initiatives Committee} from the IEEE Board of Directors and, in 2023, he has been appointed as IEEE ComSoc \textit{Distinguished Lecturer}.
\end{IEEEbiography}

\end{document}